\documentclass[11pt]{article}

\pdfoutput=1
\usepackage[T1]{fontenc}
\usepackage[latin9]{inputenc}
\usepackage[a4paper]{geometry}
\usepackage[active]{srcltx}
\usepackage{amsmath}
\usepackage{amssymb}
\usepackage{esint}
\usepackage{ulem}
\usepackage{enumitem}

\makeatletter

\usepackage{textcomp}

\pdfoutput=1 

\usepackage{jheppub}

\newcommand*\xbar[1]{%
  \hbox{%
    \vbox{%
      \hrule height 0.5pt 
      \kern0.3ex
      \hbox{%
        \kern-0.0em
        \ensuremath{#1}%
        \kern-0.0em
      }%
    }%
  }%
}

\usepackage{amsfonts}

\usepackage{stmaryrd}
\usepackage{mathtools}

\usepackage{natbib}
\usepackage{mathrsfs}
\usepackage{verbatim}

\usepackage{bbm}

\newcommand{\be}{\begin{equation}}
\newcommand{\ee}{\end{equation}}
\newcommand{\bea}{\begin{eqnarray}}
\newcommand{\eea}{\end{eqnarray}}

\title{\boldmath  Matching conditions at null infinity in the presence of logarithms: the role of advanced and retarded radiation}

\author[a]{Mat\'ias Brice\~no}
\author[a]{, Hern\'an A. Gonz\'alez}
\author[b,c]{, Marc Henneaux}
\author[d,e,f]{and Alfredo P\'erez}

\affiliation[a]{Facultad de Ingenier\'ia, Universidad San Sebasti\'an, Santiago 8420524, Chile}
\affiliation[b]{Universit\'e Libre de Bruxelles and International Solvay Institutes,  ULB-Campus Plaine CP231, B-1050 Brussels, Belgium}
\affiliation[c]{Coll\`ege de France,  Universit\'e PSL, 11 place Marcelin Berthelot,  75005 Paris, France}
\affiliation[d]{Centro de Estudios Cient\'ificos (CECs), Avenida Arturo Prat 514, Valdivia, Chile}
\affiliation[e]{Facultad de Ingenier\'ia, Universidad San Sebasti\'an, sede Valdivia, General Lagos 1163, Valdivia 5110693, Chile}
\affiliation[f]{Erwin Schr\"odinger Int.\ Institute for Mathematics and Physics, University of Vienna}

\emailAdd{mbricenoc3@correo.uss.cl}
\emailAdd{hernan.gonzalez@uss.cl}
\emailAdd{marc.henneaux@ulb.be}
\emailAdd{alfredo.perez@uss.cl}

\preprint{}

\abstract{ We provide a new perspective on the general matching conditions between the future of past null infinity and the past of future null infinity, emphasizing the impact of dominant logarithmic terms in the asymptotic expansion of the fields near null infinity.  We explicitly consider the cases of a massless scalar field and of electromagnetism.  Key in our derivation is the identification of the physical origin of these logarithms, which are associated with advanced and retarded radiation saturating the finite energy flux condition at null infinity (in a space of functions which is made precise).   The matching conditions arise then from the requirement of Coulombic (i.e., $1/r$) behaviour at spatial infinity.}

\makeatother

\begin{document}
\maketitle \flushbottom

 \newpage{}

\section{Introduction}

Matching conditions between the future of past null infinity and the past of future null infinity play a central role in the study of conservation laws and scattering processes involving massless fields \cite{Strominger:2013jfa,He:2014cra,Kapec:2015ena,Strominger:2017zoo}.  As explained in \cite{Strominger:2017zoo} (in particular text above (5.2.10)), these matching conditions rely on a certain number of assumptions concerning the decay of the fields at null infinity and their behaviour near its boundaries.
This paper investigates from a physical viewpoint the matching conditions in a more general context when these assumptions are relaxed,  both for a massless scalar field and for free electromagnetism in four dimensional Minkowski space.  

To formulate the problem in a concrete manner, let us consider the definite case of a massless scalar field, denoted by $\phi$.  The standard behaviour assumed near future null infinity is $\phi \sim \frac{a(u, x^A)}{r}$ where $a(u, x^A) =O(1)$ as $u \rightarrow - \infty$, i.e., $\lim_{u \rightarrow - \infty} a(u, x^A)= A(x^A)$ for some definite function $A(x^A)$ on the sphere.  Here, $u$ is the retarded time $t-r$ and $x^A$ are angular coordinates.  Similarly, one assumes near past null infinity (with $v$ the advanced time  $t+r$) that $\phi \sim \frac{b(v, x^A)}{r}$ with $\lim_{v \rightarrow   \infty} b(v, x^A)= B(x^A)$ for some definite function $B(x^A)$ on the sphere.  The matching conditions for the scalar field are $A(x^A) = B(-x^A)$ where $x^A \rightarrow - x^A$ stands for the antipodal map \cite{Campiglia:2017dpg,Campiglia:2017xkp}.   The matching conditions for electromagnetism and gravity take similar forms. Further matching conditions involving the subleading orders and associated with an infinite tower of charges have been investigated in 
 \cite{Briceno:2025ivl}.

 A spatial-infinity-based derivation of the matching conditions was given in \cite{Henneaux:2018gfi,Henneaux:2018hdj,Henneaux:2018mgn}, where they were shown to follow from (and in fact be equivalent with) the assumption that the leading orders of the field in the asymptotic expansion near spatial infinity should obey appropriate parity conditions under the sphere antipodal map.  These parity conditions make the action finite ``on the nose'' (without need for regularization).  In the scalar field case, the parity conditions read explicitly, in terms of the field $\phi(r, x^A)$ and its conjugate $\pi(r, x^A)$ on the initial slice $t=0$\footnote{These conditions are invariant under Poincar\'e transformations.},
\begin{equation}
\phi = \frac{C(x^A)}{r} + o(r^{-1}) \, , \qquad \pi = \frac{P(x^A)}{r^2} + o(r^{-2})  \label{Eq:ScalarCauchy01}
\end{equation}
with 
\begin{equation}
C(- x^A) = C(x^A) \, , \qquad P(-x^A) = - P(x^A) \, . \label{Eq:ScalarCauchy02}
\end{equation}

The reason that these parity conditions on the initial data are equivalent to the matching conditions of \cite{Strominger:2013jfa,He:2014cra,Kapec:2015ena,Strominger:2017zoo} can be traced to an interesting property of the asymptotic hyperbolic coordinates that connect spacelike infinity to null infinity \cite{Ashtekar:1978zz,Beig:1982ifu}, namely, that  parity properties under the sphere antipodal map on Cauchy hyperplanes become parity properties under the hyperboloid antipodal map, involving not only the sphere antipodal map but also an hyperbolic time inversion that reverses past and future \cite{Compere:2011ve,Troessaert:2017jcm}.

Now, although very general \cite{Christodoulou:1993uv}, the behaviour assumed in \cite{Strominger:2013jfa,He:2014cra,Kapec:2015ena,Strominger:2017zoo} near the past (future) of future (past) null infinity is not the most general one that is compatible with the condition of finite energy flux through null infinity.  Taking again the example of a free massless scalar field in Minkowski space, one can show that
the total energy radiated through $\mathscr{I}^{+}$ is given by \cite{Frolov:1977bp}
\begin{equation}
\left.\Delta E\right|_{\mathscr{I}^{+}}=-\int_{\mathscr{I}^{+}}dud^{2}\hat{x}\sqrt{\gamma}\left(\partial_{u}a\right)^{2}\,.\label{eq:energy_intro}
\end{equation}
Here $\gamma$ denotes the determinant of the metric of the round
2-sphere $\gamma_{AB}$. This allows a logarithmic behaviour of $a(u)$ in the limit $u \rightarrow - \infty$, which is clearly more singular than the $O(1)$-behaviour mentioned above\footnote{If we assume that in the limit $u \rightarrow - \infty$,   $\partial_u a = O(u^k)$ where $k \in \mathbb{Z}$ is an integer, then convergence holds for $k \leq - 1$.  We shall say that the upper value $k = - 1$ saturates the finite energy flux condition.  It is natural to adopt $k \in \mathbb{Z}$ because fractional powers of $u$ would yield fractional powers of $r$ at spatial infinity, which we exclude in this paper.}.

The question is then: what do the matching conditions become if one includes this more general logarithmic behaviour?  The answer to that question was given in \cite{Fuentealba:2024lll,Fuentealba:2025ekj} by following again a spatial-infinity-based route in which one integrates the field equations from initial data that keep the same standard $1/r$ behaviour of the elementary solution of the Poisson equation at spatial infinity, but with leading coefficients that do not obey  definite parity condition and contain instead  both parities (i.e., no parity restriction).  More explicitly, in the scalar field case, one keeps (\ref{Eq:ScalarCauchy01}) --  which we call ``Coulomb behaviour'' even when the coefficients $C$ and $P$ depend on the angles --,  but one drops  (\ref{Eq:ScalarCauchy02}). Such initial data were shown to lead to the above $\log (-u)/r$ behaviour near the past of future null infinity,  which is paired with a leading $\log r/r$ near null infinity.  The coefficients of these $\log (-u)/r$ and $\log r/r$ logarithmic terms come from the other parity component at spatial infinity and obey therefore opposite matching conditions to those of  \cite{Strominger:2013jfa,He:2014cra,Kapec:2015ena,Strominger:2017zoo}.  Such mixed matching conditions were found earlier in higher dimensions in \cite{Henneaux:2019yqq}.

The purpose of this paper is to provide new insight on the matching conditions through a different, physically motivated explanation for the emergence
of logarithmic terms in the asymptotic expansion at null infinity
of  massless scalar and electromagnetic fields in Minkowski space.   Our approach is complementary to that of \cite{Fuentealba:2024lll,Fuentealba:2025ekj} in that it relies from the very beginning on the behaviour of the fields near null infinity rather than on their Cauchy development from initial data.

We show that logarithmic terms at future null infinity originate from advanced waves that saturate the physically sensible condition that the total energy entering the spacetime through past null infinity is finite. This requirement yields an advanced solution exhibiting a $\log\left(v\right)$ behavior at late advanced times (rather than the stronger $O(1)$ behaviour), which, in
turn, leaves a distinct imprint at future null infinity characterized by a $\log\left(r\right)/r$ decay near $\mathscr{I}^{+}$, which dominates the usually assumed $r$-asymptotic expansion.  Analogously,
logarithmic terms at past null infinity originate from retarded waves
exhibiting a $\log\left(-u\right)$ behavior at early retarded times (see Figure 1).  The asymptotic conditions on the fields at spatial infinity, which are requested to keep the Coulombic behaviour (\ref{Eq:ScalarCauchy01}), connects the advanced and retarded branches, from which one reads the matching conditions.  These coincide with those of \cite{Fuentealba:2024lll,Fuentealba:2025ekj}, of which it provides an alternative derivation.

\begin{figure}[h]
\centering
\includegraphics[width=0.8\textwidth]{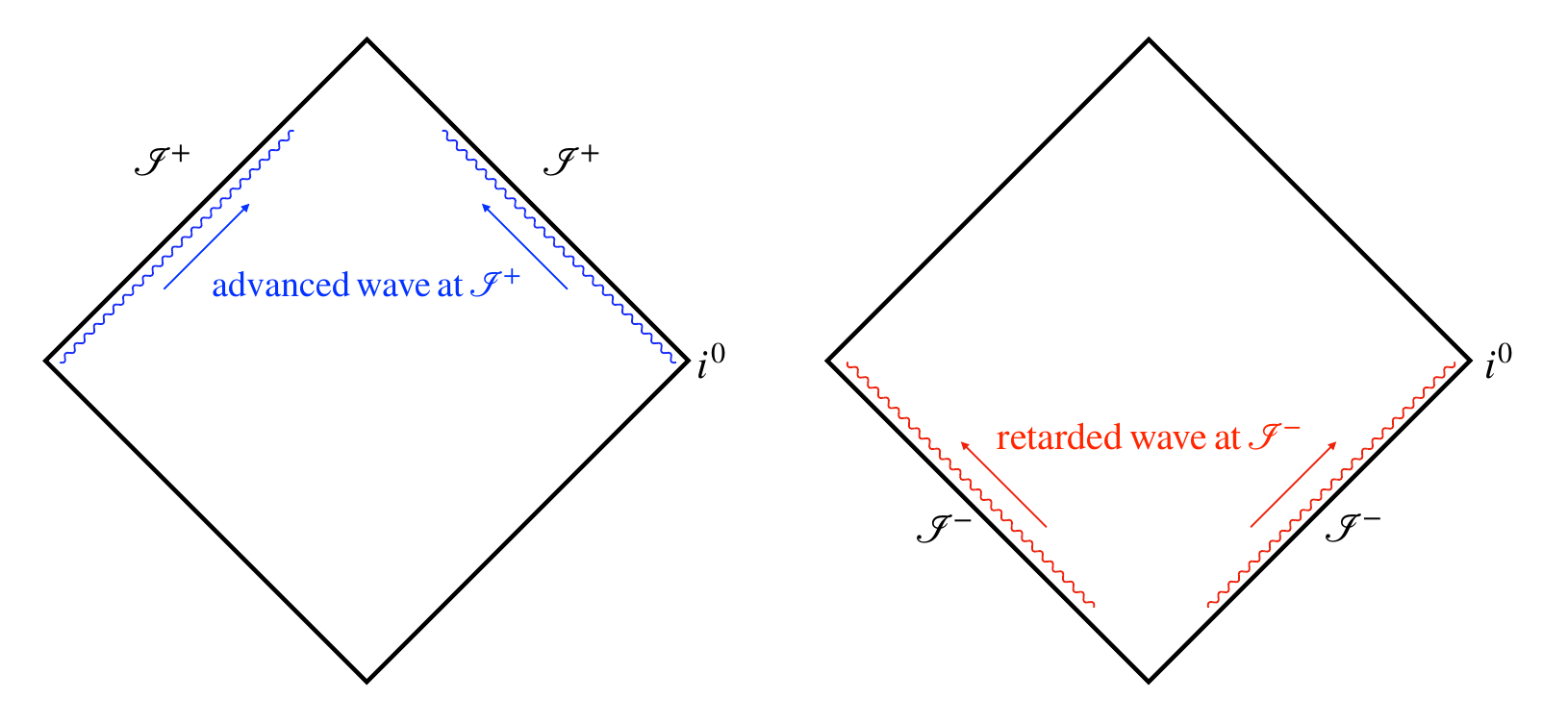}
\caption{
The double cover of the Penrose diagram for Minkowski space depicting, in blue, the trace of the last advanced modes at future null infinity $\mathscr{I}^+$ and, in red, the first retarded modes at past null infinity $\mathscr{I}^-$. 
The advanced branch corresponds to solutions whose logarithmic falloff $\log(r)/r$ at $\mathscr{I}^+$ originates from finite-energy data entering through $\mathscr{I}^-$, while the retarded branch encodes the matching conditions at early times. 
Together, they illustrate how logarithmic terms at null infinity naturally emerge from the interplay between advanced and retarded solutions saturating the finite-energy condition.
}
\end{figure}

The connection between advanced waves and non-power-law radial expansions of the fields near $\mathscr{I}^{+}$ violating the peeling property of $\mathscr{I}^{+}$ has been recognized since the earliest studies of gravitational radiation at null infinity. In fact, the standard power-law expansion of the leading orders in the asymptotic expansion in the BMS formalism might be considered to be equivalent to Sommerfeld's radiation condition, which is required to describe radiation emitted by bounded sources (see, e.g., \cite{Bondi:1962px,Sachs:1961zz,Sachs:1962wk,Tamburino:1966zz,Kroon:1998dv}). For this reason, such
terms were typically neglected due to the imposition of the Sommerfeld
radiation condition. Nonetheless, there could be physically relevant
scenarios where incoming radiation saturating the flux bound, and consequently leading logarithmic terms,
could play a crucial role. This occurs for instance in the study of the $S$-matrix where Arefeva-Faddeev-Slavnov type of boundary conditions \cite{Arefeva:1974jv,Faddeev:1980be} are relevant \cite{Kim:2023qbl,Kraus:2025wgi}.  In particular, a $\log r$ dominant term in the angular component of the vector potential involving the Feynman propagator has been explicitly exhibited in \cite{Kim:2023qbl} (appendix D, formula (D.8) and sentence below).

Recent articles \cite{Sahoo:2018lxl,Campiglia:2019wxe,AtulBhatkar:2019vcb,AtulBhatkar:2021txo,Compere:2025tzr} have also addressed the problem of matching conditions in the presence of logarithms, confirming the ``mixed matching conditions phenomenon" obtained by integrating the fields from initial data given on Cauchy hypersurfaces.  However, these interesting papers do not treat the leading logarithms that motivate the present work.  Of course,  one recovers the conditions studied in these papers as a special case  if one sets the coefficients of the leading logarithms equal to zero. 

Our paper is organized as follows.  In the next section (Section {\bf \ref{Sec:Scalar}}), we treat the case of the scalar field.  We cover in detail monopole and dipole radiation fields, which exhibit the main points.   We then move in Section {\bf \ref{Sec:EM}} to the electromagnetic case, of which we directly provide the all-multipole analysis. We first give the  form of the electromagnetic potential and of the electromagnetic field.  We investigate next the conditions under which the energy fluxes through past and future infinity are finite. These conditions, together with  the requested power law behaviour at spatial infinity, lead then to the matching conditions for electromagnetism. Section {\bf \ref{Sec:Conclusions}} is devoted to conclusions and prospects.  Three appendices complete our paper: Appendix {\bf \ref{App:Gen}} covers the case of a scalar field with a general multipole expansion. Appendix {\bf \ref{App:Max}} provides a thorough derivation of the general solution of Maxwell's equations in retarded coordinates.  Finally, Appendix  {\bf  \ref{LegendreAppendix}} lists useful identities on infinite sums associated with the Legendre polynomials.

\section{Massless scalar field}
\label{Sec:Scalar}
We start with a massless scalar field in four spacetime dimensions, obeying the wave equation
\begin{equation}
\Box \phi = 0\, ,
\end{equation}
which reads explicitly, in retarded coordinates $(u= t-r, r, x^A)$,
\begin{equation}
-2 \partial_u \partial_r \phi  + \partial_r^2 \phi  + \frac{2 \partial_r \phi}{r} - \frac{2 \partial_u \phi}{r} + \frac{D^2 \phi}{r^2} = 0\,  \label{Eq:uWaveEq}
\end{equation}
where $D^2$ is the Laplacian on the unit $2$-sphere.

  We impose the following asymptotic conditions:
\begin{itemize}
\item The field can be written as the sum of retarded and advanced waves
\begin{equation}
\phi=\frac{\bar{\phi}_{R}\left(u, x^A\right)}{r}+\frac{\bar{\phi}_{A}\left(v, x^A\right)}{r}+ O (r^{-2})\,,\label{eq:phi_solR+A0}
\end{equation}
where the functions $\bar{\phi}_{R}$ and $\bar{\phi}_{A}$, which are smooth but not necessarily analytic in $u = t-r$ or $v=t+r$,  are required to give finite energy fluxes through null infinity.  The fact that we write the field $\phi$ as a sum of a retarded part and an advanced one does not mean, of course, that $u$ and $v$ are independent variables (given $r$), but simply that there is one part of $\phi$ that is a function of time through $u$ and the other, through $v$. When computing the wave equation in retarded coordinates (say), one should of course replace $v$ by $u+2r$ (and vice-versa, use $u = v-2r$ when working in advanced coordinates).
\item The field $\phi$ has the ``Coulombic'' $1/r$ behaviour at spatial infinity, characteristic of the elementary solution of Poisson's equation, i.e., on constant Minkowskian time hyperplanes,
\begin{equation}
\phi=\frac{C(x^A)}{r}+o (r^{-1})\,,\label{eq:phi_solCoulomb}
\end{equation}
and its conjugate momentum behaves as
\begin{equation}
 \pi = \frac{P(x^A)}{r^2} + o(r^{-2}) \, ,\label{eq:phi_solCoulombConjugate}
\end{equation}
in agreement with (\ref{Eq:ScalarCauchy01}). But note that coefficients that depend on the angles are allowed.
\end{itemize}
\subsection{Spherical waves}

\subsubsection{Logarithmic terms at null infinity from advanced and retarded radiation}

We first illustrate the approach by considering the simplest case
of a spherical wave. In this case, the functions $\bar{\phi}_{R}$ and $\bar{\phi}_{A}$ do not depend on the angles and the general solution to the wave
equation reduces to\footnote{Note that the split of the static component $\phi = \frac{a}{r}$ is ambiguous since it corresponds to $\bar{\phi}_{R}\left(u\right) = a - \lambda$, $\bar{\phi}_{A}\left(v\right) =  \lambda$, with $\lambda$ arbitrary.  We will lift this ambiguity by imposing that the constant $a$ be equally split between the retarded and advanced parts, i.e., $\lambda = \frac{a}{2}$.}
\begin{equation}
\phi=\frac{\bar{\phi}_{R}\left(u\right)}{r}+\frac{\bar{\phi}_{A}\left(v\right)}{r}\,.\label{eq:phi_sol-2}
\end{equation}
It is important to emphasize that the functions $\bar{\phi}_{R}\left(u\right)$
and $\bar{\phi}_{A}\left(v\right)$ are not determined by the differential
equation and are completely arbitrary. Nonetheless, this
freedom must be partially constrained by the fundamental physical
requirement that the total radiated energy be finite.

As indicated above, an expression for the total radiated
or absorbed energy was derived in Ref. \cite{Frolov:1977bp}, by projecting the energy-momentum tensor
of the scalar field onto $\mathscr{I}^{\pm}$. In particular, for
the spherically symmetric solution given in Eq.~\eqref{eq:phi_sol-2},
the total energy radiated through $\mathscr{I}^{+}$ by the retarded field is given by
\begin{equation}
\left.\Delta E\right|_{\mathscr{I}^{+}}=-\int_{\mathscr{I}^{+}}dud^{2}\hat{x}\sqrt{\gamma}\left(\partial_{u}\bar{\phi}_{R}\right)^{2}\,.\label{eq:energy_ret}
\end{equation}
To ensure that the total energy radiated by the retarded field
remains finite, in the limit $u\rightarrow\pm\infty$, the behavior
of the field $\bar{\phi}_{R}$ near $\mathscr{I}_{+}^{+}$ and $\mathscr{I}_{-}^{+}$
must be, at most, of the form  (in the space of functions such that $\partial_u f \sim u^k$, $k \in \mathbb{Z}$):
\begin{equation}
\bar{\phi}_{R}\left(u\right)\underset{u\rightarrow\pm\infty}{=}\phi_{R}^{\pm}\log\left(\pm u\right)+\varphi_{R}^{\pm}+o(1),\label{eq:phiR-1}
\end{equation}
where the coefficients $\phi_{R}^{\pm}$ and $\varphi_{R}^{\pm}$
are constants for the spherical wave. 

Analogously, the total radiation entering the spacetime through $\mathscr{I}^{-}$
is 
\begin{equation}
\left.\Delta E\right|_{\mathscr{I}^{-}}=\int_{\mathscr{I}^{-}}dvd^{2}\hat{x}\sqrt{\gamma}\left(\partial_{v}\bar{\phi}_{A}\right)^{2}\,.\label{eq:energy_adv}
\end{equation}
Therefore, the condition for having a finite total incoming energy
is
\begin{equation}
\bar{\phi}_{A}\left(v\right)\underset{v\rightarrow\pm\infty}{=}\phi_{A}^{\pm}\log\left(\pm v\right)+\varphi_{A}^{\pm}+ o(1) \,,\label{eq:phiA-1}
\end{equation}
where $\phi_{A}^{\pm}$ and $\varphi_{A}^{\pm}$ are constants for
the spherical wave. 

The key step is to rewrite Eq.~\eqref{eq:phi_sol-2} entirely in terms
of the retarded time, $u$ using the relation $v=u+2r$. In this form,
the general solution for a spherically symmetric wave becomes:
\begin{equation}
\phi=\frac{\bar{\phi}_{R}\left(u\right)}{r}+\frac{\bar{\phi}_{A}\left(u+2r\right)}{r}.\label{eq:phi_sol-1-2}
\end{equation}
Taking the limit to future null infinity ($u=\text{const}$ and $r\rightarrow\infty$),
the argument of the advanced component of the solution is dominated
by $r$. Thus, using the asymptotic behavior (\ref{eq:phiA-1}), one
finds
\begin{equation}
\left.\phi\right|_{\mathscr{I}^{+}}=\phi_{A}^{+}\frac{\log r}{r}+\left(\bar{\phi}_{R}\left(u\right)+\phi{}_{A}^{+}\log\left(2\right)+\varphi_{A}^{+}\right)\frac{1}{r}+o(r^{-1}) \,,\label{eq:phi_sol-1-1-1}
\end{equation}
which is precisely the logarithmic behavior described in \cite{Fuentealba:2024lll}
\begin{equation}
\left.\phi\right|_{\mathscr{I}^{+}}=\frac{1}{2}\Psi\frac{\log\left(r\right)}{r}+\frac{\Phi}{r}+o(r^{-1})\,,\label{eq:psi}
\end{equation}
with 
\begin{equation}
\Phi=\bar{\phi}_{R}\left(u\right)+\phi{}_{A}^{+}\log\left(2\right)+\varphi_{A}^{+}\,,\label{eq:Phi}
\end{equation}
and 
\[
\Psi=2\phi_{A}^{+}.
\]

This analysis shows that the late-time behavior of the advanced wave
manifests itself at future null infinity as a term of the form $\log\left(r\right)/r$. One might fear that this $\frac{\log\left(r\right)}{r}$-term will jeopardize the finite energy flux condition due to its slowlier decay at null infinity, but this is not the case because it does not depend on $u$ so that its $u$-derivative vanishes. 
Note also that the term of order $r^{-1}$ acquires a time-independent
component from the advanced solution, which, due to its independence
of $u$ , is not part of the radiative field at future null infinity.  The other $\frac{1}{r}$-contributions from the advanced wave depend on $u$, but are subleading in the limit $u \rightarrow- \infty$ (decay as $\sim u^{-1}$ or faster \cite{Fuentealba:2024lll}).

Analogously, if we perform a similar analysis at past null infinity
$\mathscr{I}^{-}$ one finds
\[
\left.\phi\right|_{\mathscr{I}^{-}}=\phi_{R}^{-}\frac{\log r}{r}+\left(\bar{\phi}_{A}\left(v\right)+\phi_{R}^{-}\log\left(2\right)+\varphi_{R}^{-}\right)\frac{1}{r}+o(r^{-1})\,.
\]
Therefore, in the expansion at past null infinity corresponding to (\ref{eq:psi})
\begin{equation}
\left.\phi\right|_{\mathscr{I}^{-}}=\frac{1}{2}\Psi'\frac{\log\left(r\right)}{r}+\frac{\Phi'}{r}+o(r^{-1})\,,\label{eq:psiprime}
\end{equation}
one identifies
\begin{align}
\Phi' & =\bar{\phi}_{A}\left(v\right)+\phi_{R}^{-}\log\left(2\right)+\varphi_{R}^{-}\,,\label{eq:phiprime-1}
\end{align}
and
\[
\Psi'=2\phi_{R}^{-}.
\]

\subsubsection{Matching conditions}
We now derive the matching conditions for spherical
waves near spatial infinity $i^{0}$. 

For the retarded solution, spatial
infinity is approached in the limit $u\rightarrow-\infty$ with large
$r$, such that the combination $t=u+r$ remains finite. Using Eq.~\eqref{eq:phiR-1} one finds
\begin{align*}
\lim_{u\rightarrow-\infty}\frac{\bar{\phi}_{R}}{r} & =\lim_{r\rightarrow+\infty}\frac{\phi_{R}^{-}\log\left(r-t\right)}{r}+\frac{\varphi_{R}^{-}}{r}=\phi_{R}^{-}\frac{\log\left(r\right)}{r}+\frac{\varphi_{R}^{-}}{r}+\dots\,,
\end{align*}
Analogously, the limit of the advanced solution to spatial infinity
is given by the limit $v\rightarrow+\infty$ with large $r$, such
that the combination $t=v-r$ remains finite. Therefore, using Eq.~\eqref{eq:phiA-1} one finds
\begin{align*}
\lim_{v\rightarrow+\infty}\frac{\bar{\phi}_{A}}{r} & =\lim_{v\rightarrow+\infty}\frac{\phi_{A}^{+}\log\left(r+t\right)}{r}+\frac{\varphi_{A}^{+}}{r}=\frac{\phi_{A}^{+}\log\left(r\right)}{r}+\frac{\varphi_{A}^{+}}{r}+\dots\,.
\end{align*}

To obtain the full solution at $i^{0}$, we must consider the sum
of the contributions from the advanced and retarded waves. Consequently,
the leading order of the solution near $i^{0}$ is given by
\[
\left.\phi\right|_{i^{0}}=\left(\phi_{R}^{-}+\phi_{A}^{+}\right)\frac{\log\left(r\right)}{r}+\frac{\varphi_{A}^{+}+\varphi_{R}^{-}}{r}+\dots\,.
\]

As recalled in the introduction and argued by many authors, the natural behaviour at spatial infinity is the one of the elementary solution of Poisson equation, i.e., $\phi \sim \frac1r$. 
Thus,  
the $\log r$ terms should cancel in the expansion of the field near $i^{0}$.
This implies that one must impose the following condition: 
\begin{equation}
\phi_{R}^{-}=-\phi_{A}^{+}.\label{eq:matchlog-1}
\end{equation}
In terms of the variables defined in Eqs. (\ref{eq:psi}) and (\ref{eq:psiprime}),
this becomes
\[
\Psi=-\Psi',
\]
which precisely coincides with the matching condition established
in \cite{Fuentealba:2024lll} for the coefficient of the leading logarithmic $\log  r/r$ term, applied to the spherical wave.  This matching condition corresponds to the $Q$ branch in the language of \cite{Fuentealba:2024lll} and fulfills matching conditions with the minus sign, opposite to the matching condition adopted in \cite{Strominger:2013jfa,He:2014cra,Kapec:2015ena,Strominger:2017zoo} and valid for the subleading $P$-branch. 

It is through the matching conditions that the coefficients of the $\frac{\log r}{r}$-term and of the  $\frac{\log u}{r}$-term in the expansion near future null infinity are forced to be equal (up to the sign).  They would otherwise be unrelated, having a priori different origins.

The $P$-branch is the dominant one if one assumes from the outset that there is no leading $\log r/r$ term in the expansion of the scalar field near future null infinity, i.e., $\phi_{A}^{+} = 0$ (and thus also $\phi_{R}^{-} = 0$ by (\ref{eq:matchlog-1})).  In that case one finds that $\bar{\phi}_{R}\left(u\right)$ (respectively, $\bar{\phi}_{A}\left(v\right)$) behaves near the past of future null infinity (respectively, the future of past null infinity) as
\begin{equation}
\bar{\phi}_{R}\left(u\right) = \varphi_{R}^{-} \, , \qquad \bar{\phi}_{A}\left(v\right) = \varphi_{A}^{+}\label{eq:PSpherical}
\end{equation}
up to terms that vanish in the limit.  Hence, one has
\begin{equation}
\Psi = \Psi' = 0 \, , \qquad  \lim_{u\rightarrow-\infty} \Phi=\varphi_{R}^{-}+\varphi_{A}^{+} = + \lim_{v\rightarrow \infty} \Phi' \, ,  \label{eq:matchP-1}
\end{equation}
(and $\left.\phi\right|_{i^{0}}=\frac{\varphi_{A}^{+}+\varphi_{R}^{-}}{r}+\dots$) which is the matching condition, with a positive sign, for the $P$-branch \cite{Campiglia:2017dpg}. It is of interest to emphasize that the matching conditions, which involve the sum $\varphi_{A}^{+}+\varphi_{R}^{-}$, are insensitive to the ambiguity in the decomposition of the static Coulomb part between retarded and advanced solutions.  

As explained in \cite{Fuentealba:2024lll}, when the $Q$-branch is present ($\Psi \not=0$,  $\Psi' \not=0$), one must substract its contributions to the $1/r$-term in order to dig out the $P$-contributions, which obey (\ref{eq:matchP-1}).

We stress that the retarded and advanced solutions, when expressed in terms of the retarded or advanced coordinate, have a purely
power-law behaviour in the radial coordinate (in this case just $1/r$), with no logarithmic
term involving $r$.  The logarithms at future null infinity appear only when the advanced solution is expressed in terms of the retarded time (and similarly for past null infinity).  For that reason, the logarithms at future null infinity can be eliminated by assuming no advanced radiation. By the matching at spatial infinity, this would also eliminate the $\log (-u)$ term in $\bar \phi_R$.

\subsection{Dipole waves}

\subsubsection{Logarithmic terms at null infinity from advanced and retarded radiation}

Since the equations become increasingly involved when one includes higher spherical harmonics, we shall treat explicitly the dipole case in the core of the text.  It provides a very good illustration of the general case, which is covered in Appendix {\bf \ref{App:Gen}}.

The general solution of the wave equation in a source-free region
can again be expressed as the sum of retarded and advanced terms.
\[
\phi=\phi_{R}+\phi_{A},
\]
where we have now, following \cite{Bondi:1962px}, 
\begin{align}
\phi_{R} & =\sum_{m= -1, 0, 1}\left(\frac{1}{r}\frac{d a_m\left(u\right)}{du}- \frac{1}{r^2}a_m\left(u\right)\right) Y_{1 m}\left(\theta,\phi\right),\label{eq:phi_R07}\\
\phi_{A} & =\sum_{m= -1, 0, 1}\left(\frac{1}{r}\frac{d b_m\left(v\right)}{dv}- \frac{1}{r^2}b_m\left(v\right)\right) Y_{1 m}\left(\theta,\phi\right).\label{eq:phi_A07}
\end{align}
Here, $a_{m}\left(u\right)$ and $b_{m}\left(v\right)$
are arbitrary functions of the retarded and advanced times, respectively.  We find again the same ambiguity in the decomposition between retarded and advanced parts of  the Coulomb static component, now at dipole order $1/r^2$. 

Consider first the retarded solution.  The total energy flux across $\mathscr{I}^{+}$ takes exactly the same form as in Eqs. (\ref{eq:energy_ret}) where $\bar{\phi}_{R}\left(u,\hat{x}\right)$ is the coefficient of the leading $1/r$ expansion,
\begin{equation}
\phi_R=\frac{\bar{\phi}_{R}\left(u,\hat{x}\right)}{r}+O\left(r^{-2}\right),
\end{equation}
with
\begin{equation}
\bar{\phi}_{R}\left(u,\hat{x}\right)  =\sum_{m}\frac{d a_{m}\left(u\right)}{du}\, Y_{1 m} \, .
\end{equation}
Finiteness of the total energy flux implies therefore the condition
\begin{equation}
\lim_{u\rightarrow\pm\infty}\frac{d a_{m}\left(u\right)}{du} =\phi_{R\left(m\right)}^{\pm}\log\left(\pm u\right)+\varphi_{R\left(m\right)}^{\pm}+\dots\,,
\end{equation}
which leads upon integration, near the past of future null infinity ($u \rightarrow - \infty$) to
\begin{equation}
a_{m}\left(u\right) =\phi_{R\left(m\right)}^{-} \left( u \log (-u) - u \right) + \varphi_{R\left(m\right)}^{-} u + c_{R\left(m\right)} + \dots 
\end{equation}
where the dots denote subleading terms (in $u$).
Similarly, the condition of finiteness of the total flux through past infinity implies ($v \rightarrow  \infty$)
\begin{equation}
b_{m}\left(v\right) =\phi_{A\left(m\right)}^{+} \left( v \log (v) - v \right) + \varphi_{A\left(m\right)}^{+} v + c_{A\left(m\right)} + \dots  \label{eq:db07Dip}
\end{equation}

Substituting Eq.~\eqref{eq:db07Dip} into Eq.~\eqref{eq:phi_A07}, replacing $v$ by $u+2r$
and taking the limit $r\rightarrow\infty$ with $u$ held constant (which implies $v\rightarrow\infty$), one finds that the solution can then be written
 near $\mathscr{I}^{+}$ as:
\begin{align}
\left.\phi\right|_{\mathscr{I}^{+}} & =- \left[\sum_{m}\phi_{A\left(m\right)}^{+}Y_{l m}\right]\frac{\log r}{r}\nonumber \\
 & +\left[\sum_{m}\left(\frac{d}{du}a_{m}\left(u\right)-\left(\phi_{A\left(m\right)}^{+}\log2+\varphi_{A\left(m\right)}^+\right)+2\phi_{A\left(m\right)}^{+}\right)Y_{l m}\right]\frac{1}{r}+o(r^{-1})\,.\label{eq:phi_I+07}
\end{align}
Note again that a $\log r/r$ term appears, arising exclusively from the
advanced wave. This term, which comes both from the $1/r$ and the $1/r^2$ contributions to $\phi_A$, is time-independent and thus does not contribute to the energy flux  across $\mathscr{I}^{+}$. Additionally,
the advanced wave also contributes a time-independent term at order
$r^{-1}$, arising again from the $1/r$ and the $1/r^2$ terms in $\phi_A$. The $\log(- u)$ term (if any) is contained in the retarded component and is at this stage unrelated to the $\log r/r$ term.

Let us now determine the
asymptotic expansion near $\mathscr{I}^{-}$. By similar manipulations, one finds that the full solution reads
\begin{align}
\left.\phi\right|_{\mathscr{I}^-} & =- \left[\sum_{m}\phi_{R\left(m\right)}^{-}Y_{l m}\right]\frac{\log r}{r}\nonumber \\
 & +\left[\sum_{m}\left(\frac{d}{dv}b_{m}\left(v\right)-\left(\phi_{R\left(m\right)}^{-}\log2+\varphi_{R\left(m\right)}^-\right)+2\phi_{R\left(m\right)}^{-}\right)Y_{l m}\right]\frac{1}{r}+o(r^{-1})\,.\label{eq:phi_I-07}
\end{align}
In this case, there is a time-independent $\log r/r$ contribution
arising exclusively from the retarded wave.

The expressions (\ref{eq:phi_I+07}) and (\ref{eq:phi_I-07}), which describe
the behavior of the field near $\mathscr{I}^{\pm}$, match precisely
the asymptotic expansion introduced in Ref. \cite{Fuentealba:2024lll}
and given in Eqs. (\ref{eq:psi}) and (\ref{eq:psiprime}), where
\begin{align}
\Psi\left(\hat{x}\right) & =-2\sum_{m}\phi_{A\left(m\right)}^{+}Y_{1 m}\left(\hat{x}\right)\,,\label{eq:PsiellDip}\\
\Psi'\left(\hat{x}\right) & =-2\sum_{m}\phi_{R\left(m\right)}^{-}Y_{1 m}\left(\hat{x}\right)\,.\label{eq:psipellDip}
\end{align}
In the next subsection, we will show that this prescription directly
yields the matching conditions for the scalar field near spatial infinity.

\subsubsection{Matching conditions}

Let us examine the behavior of the field $\phi$ and its canonical
momentum $\pi=\partial_{t}\phi$ near spatial infinity $i^{0}$. 

For the advanced solution, one must replace $v=t+r$ in the solution
and then take the limit $r\rightarrow\infty$ while keeping $t$ fixed.  This gives
\begin{align*}
\left.\phi_{A}\right|_{i^{0}} & =\sum_{m}\frac{1}{r}\left[\phi_{A\left(m\right)}^{+}\left(\log\left(t+r\right)\right)+\varphi_{A\left(m\right)}^+\right]Y_{1 m} \\
& \qquad \qquad - \sum_{m}\frac{1}{r^{2}}\left[\phi_{A\left(m\right)}^{+}\left(\left(t+r\right)\log\left(t+r\right)-\left(t+r\right)\right)+\varphi_{A\left(m\right)}^+\left(t+r\right)\right]Y_{1 m}+\dots\,,\\
 & =\left[\sum_{m}\phi_{A\left(m\right)}^{+}Y_{1 m}\right]\frac{1}{r}+\dots\,.
\end{align*}
Note that the $\frac{\log r}{r}$ term drops.

Similarly, for the retarded solution one replaces $u=t-r$ in the solution and then takes the limit $r\rightarrow\infty$
while maintaining $t$ fixed.  One gets this time
\begin{equation*}
\left.\phi_{R}\right|_{i^{0}} =\left[\sum_{m}\phi_{R\left(m\right)}^{-}Y_{1 m}\right]\frac{1}{r}+\dots\, ,
\end{equation*}
again with no $\frac{\log r}{r}$ term.

Adding the two contributions, we get
\begin{equation}
\left.\phi\right|_{i^{0}} =\left[\sum_{m}\left(\phi_{A\left(m\right)}^{+} + \phi_{R\left(m\right)}^{-}\right)Y_{1 m}\right]\frac{1}{r}+\dots\, ,
\end{equation}

Let us now compute the canonical momentum $\pi=\partial_{t}\phi$
near $i^{0}$. The contribution from the advanced solution is given
by
\begin{align*}
\left.\pi_{A}\right|_{i^{0}} & =\partial_t\sum_{m}\frac{1}{r}\left[\phi_{A\left(m\right)}^{+}\left(\log\left(t+r\right)\right)+\varphi_{A\left(m\right)}^+\right]Y_{1 m} \\
& \qquad \qquad - \partial_t \sum_{m}\frac{1}{r^{2}}\left[\phi_{A\left(m\right)}^{+}\left(\left(t+r\right)\log\left(t+r\right)-\left(t+r\right)\right)+\varphi_{A\left(m\right)}^+\left(t+r\right)\right]Y_{1 m}+\dots\,,\\
 & =- \left[\sum_{m}\phi_{A\left(m\right)}^{+}Y_{1 m}\right]\frac{\log r}{r^{2}}
 +\left[\sum_{m}\left(\phi_{A\left(m\right)}^{+}-\varphi_{A\left(m\right)}^+\right)Y_{1 m}\right]\frac{1}{r^{2}}+\dots\,.
\end{align*}
Similarly, the retarded component of the momentum near $i^{0}$ takes the form
\begin{align*}
\left.\pi_{R}\right|_{i^{0}}& =\partial_t\sum_{m}\frac{1}{r}\left[\phi_{R\left(m\right)}^{-}\left(\log\left(r-t\right)\right)+\varphi_{R\left(m\right)}^-\right]Y_{1 m} \\
& \qquad \qquad + \partial_t \sum_{m}\frac{1}{r^{2}}\left[\phi_{R\left(m\right)}^{-}\left(\left(t-r\right)\log\left(r-t\right)-\left(t-r\right)\right)+\varphi_{R\left(m\right)}^-\left(t-r\right)\right]Y_{1 m}+\dots\,,\\
 & = \left[\sum_{m}\phi_{R\left(m\right)}^{-}Y_{1 m}\right]\frac{\log r}{r^{2}}
 +\left[\sum_{m}\left(-\phi_{R\left(m\right)}^{-}+\varphi_{R\left(m\right)}^-\right)Y_{1 m}\right]\frac{1}{r^{2}}+\dots\,.
\end{align*}

Thus, the combined contribution of the retarded and advanced components
of the momentum near $i^{0}$ becomes 
\begin{align}
\left.\pi\right|_{i^{0}} & =- \left[\sum_{m}\left(\phi_{A\left(m\right)}^{+}-\phi_{R\left(m\right)}^{-}\right)Y_{1 m}\right]\frac{\log r}{r^{2}}\nonumber \\
 & -\left[\sum_{m}\left(\varphi_{A\left(m\right)}^+-\varphi_{R\left(m\right)}^{-}
  +\phi_{R\left(m\right)}^{-}-\phi_{A\left(m\right)}^{+}\right)Y_{l m}\right]\frac{1}{r^{2}}+\dots\, \label{Eq:PiDipSpatial}
\end{align}

As expressed in Eqs. (\ref{eq:phi_solCoulomb}) and 
(\ref{eq:phi_solCoulombConjugate}), terms of order $\log r/r$
in $\left.\phi\right|_{i^{0}}$ and $\log r/r^{2}$ in $\left.\pi\right|_{i^{0}}$
are not allowed. Therefore, from Eq.~\eqref{Eq:PiDipSpatial},
we obtain the following condition
 between the coefficients
$\phi_{A\left(m\right)}^{+}$ and $\phi_{R\left(m\right)}^{-}$
\begin{equation}
\phi_{A\left(m\right)}^{+}=\phi_{R\left(m\right)}^{-}.\label{eq:phiAR00}
\end{equation}

The above expression establishes a relation between the leading logarithmic
contributions of the field at $\mathscr{I}_{-}^{+}$ and $\mathscr{I}_{+}^{-}$.
Using the definitions of the radial logarithmic terms given in Eqs.(\ref{eq:PsiellDip})
and (\ref{eq:psipellDip}), and imposing the condition \eqref{eq:phiAR00},
one obtains:
\begin{align*}
\Psi\left(\hat{x}\right)+\Psi'\left(-\hat{x}\right) & =-2\sum_{m}\left(\phi_{A\left(m\right)}^{+}Y_{1m}\left(\hat{x}\right)+\phi_{R\left(m\right)}^{-}Y_{1 m}\left(-\hat{x}\right)\right)\,,\\
 & =-2\sum_{m}\left(\phi_{A\left(m\right)}^{+}-\phi_{R\left(m\right)}^{-}\right)Y_{1 m}\left(\hat{x}\right)\,,\\
 & =0\,
\end{align*}
(using the parity property of the spherical harmonics,
$Y_{1 m}\left(-\hat{x}\right)=-Y_{1 m}\left(\hat{x}\right)$), 
which is precisely   the matching condition 
\[
\Psi\left(\hat{x}\right)=-\Psi'\left(-\hat{x}\right),
\]
established
in \cite{Fuentealba:2024lll}.

If we assume that there is no leading logarithmic term at null infinity, i.e., $\phi_{A\left(m\right)}^{+}=0 = \phi_{R\left(m\right)}^{-}$, then the next $1/r$ term (which becomes leading) reads
\begin{align}
\left.\phi\right|_{\mathscr{I}^{+}} & = \frac{\Phi\left(\hat{x}\right)}{r} +o(r^{-1}) \, , \qquad  \Phi\left(\hat{x}\right) =
 \sum_{m}\left(\varphi_{R\left(m\right)}^- -\varphi_{A\left(m\right)}^+\right)Y_{1 m}\,,\label{eq:phi_I+07bis} \\
\left.\phi\right|_{\mathscr{I}^-} & =\frac{\Phi'\left(\hat{x}\right)}{r}+o(r^{-1}) \, , \qquad  \Phi'\left(\hat{x}\right) = \sum_{m}\left(\varphi_{A\left(m\right)}^+ -\varphi_{R\left(m\right)}^-\right)Y_{1 m}\,,\label{eq:phi_I-07bis}
\end{align}
from which one infers the standard matching condition \cite{Campiglia:2017dpg}
\[
\Phi\left(\hat{x}\right)= \Phi'\left(-\hat{x}\right) \,.
\]
At the same time, the coefficient of the leading  ($1/r$) term in the expansion of $\phi$ near spatial infinity vanishes while the coefficient of the leading  ($1/r^2$) term in the expansion of $\pi$ is arbitrary, in agreement with the parity conditions recalled in the introduction (only even (respectively, odd) spherical harmonics for the leading term of $\phi$ (respectively, $\pi$)).

\subsection{General case}

The general case is treated exactly along the same lines as the monopole and dipole terms.  As the formulas become rather involved, it is dealt with in detail in Appendix {\bf \ref{App:Gen}}.  One finds in particular that the condition for the absence at spatial infinity of $\log r/r$ terms in $\phi$ ($l$ even) and $\log r/r^2$ terms in $\pi$ ($l$ odd) yields the generalized matching conditions of \cite{Fuentealba:2024lll}.

One way to understand the reason that it is the asymptotic behaviour at $i^0$ of $\phi$ for $l$ even and that of $\pi$ for $l$ odd that give non trivial conditions goes as follows.
Assume that one adopts initial values of $\phi$ and $\pi$ that violate the asymptotic decay (\ref{eq:phi_solCoulomb})-(\ref{eq:phi_solCoulombConjugate}) at spatial infinity by terms of the form originating from the logarithms at null infinty encountered previously.  That is, consider an asymptotic behaviour that reads 
$$ \phi\vert_{t=0} \sim \lambda(x^A)\frac{ \log r}{r} + \textrm{allowed terms} \, , \quad 
\pi\vert_{t=0} \sim \mu(x^A)\frac{ \log r}{r^2} + \textrm{allowed terms} \, ,$$
where the coefficients $\lambda(x^A)$  and $\mu(x^A)$ are a priori arbitrary.  One can easily  integrate the Klein-Gordon equation for such initial data in hyperbolic coordinates, all the way to null infinity, by following the method of \cite{Henneaux:2018mgn,Fuentealba:2024lll}.  The differential equation that controls the time development of $\lambda(x^A)$ and $\mu(x^A)$ can be reduced to the Legendre differential equation.  Its solution can be split into a sum of (i) a $P$-branch, which is polynomial and determined by the even part of $\lambda$ and the odd part of $\pi$; and (ii) a $Q$-branch,  which exhibits a logarithmic behaviour near null infinity and which is determined by the odd part of $\lambda$ and the even part of $\mu$.  The $Q$-branch is absent if and only if $\lambda^{\textrm{odd}}= 0$ and $\mu^{\textrm{even}} = 0$. 

By integrating the equations from the initial slice to null infinity, one then finds that for arbitrary $\lambda(x^A)$ and $\mu(x^A)$ fulfilling no particular parity properties,  the field $\phi$  takes near null infinity the form 
\begin{equation*}
\phi \sim \Theta^Q \, \frac{( \log r)^2}{r} + \textrm{subleading terms}\,,
\end{equation*}
where the coefficient $\Theta^Q$ vanishes if and only if the $Q$-branch is absent.  In $ \frac{( \log r)^2}{r}$, one logarithm originates from the assumed decay at spatial infinity ($\sim \frac{ \log r}{r}$) while the other logarithm originates from the fact that the $Q$-branch develops a logarithmic behaviour near null infinity.  The $P$-branch maintains the $ \frac{ \log r}{r}$ behaviour.

Now, the  $\frac{( \log r)^2}{r}$-term is incompatible with the assumptions made at null infinity, which only allowed  $ \frac{ \log r}{r}$-terms.  From what we have seen, this means that our assumptions at null infinity automatically enforce $\lambda^{\textrm{odd}}= 0$ and $\mu^{\textrm{even}} = 0$.  The only non-trivial components of $\lambda(x^A)$ and $\mu(x^A)$ present in our context are therefore $\lambda^{\textrm{even}}$ and $\mu^{\textrm{odd}} $.  The requirement that the initial conditions have no $ \frac{ \log r}{r}$-term for $\phi$ and $ \frac{ \log r}{r^2}$-term for $\pi$ reduces accordingly to the non-trivial conditions $\lambda^{\textrm{even}}= 0$ and $\mu^{\textrm{odd}}=0 $ since $\lambda^{\textrm{odd}}= 0$ and $\mu^{\textrm{even}} = 0$ are automatically implemented\footnote{Recall that with initial data that take the form (\ref{eq:phi_solCoulomb})-(\ref{eq:phi_solCoulombConjugate}), with no $\log$'s, the $ \frac{ \log r}{r}$-term at null infinity originates from the $Q$-branch of the $\frac{1}{r}$-term of the initial data \cite{Henneaux:2018mgn,Fuentealba:2024lll}.}.

If we restrict our attention to a definite multipole type, $\lambda$ will thus be identically zero for $l$ odd.  Requiring it to vanish will thus yield no non trivial condition. Similarly, $\mu$ will be identically zero for $l$ even so that requiring it to vanish will yield no non trivial condition.

Whatever the parity of the spherical harmonics that bring the non trivial matching conditions, we see that these originate from requirements on the asymptotic behaviour at $i^0$.  Without these requirements, the advanced and retarded parts of the scalar field would be completely independent.

\section{Electromagnetism}
\label{Sec:EM}

\subsection{The electromagnetic potential}

The previous analysis can be naturally extended to the case of the
electromagnetic field. As in the scalar case, the last advanced wave compatible with finite incoming energy leaves a $\log r$ imprint in the field strength at $\mathscr{I}^{+}$, while the first retarded wave produces a similar imprint at $\mathscr{I}^{-}$. Consistency with the standard fall-off near $i^{0}$ then requires the matching conditions derived in \cite{Fuentealba:2025ekj}.

In this section, we consider the general case that includes all possible multipole moments. As a first step, one must determine the general solution to the free Maxwell equations in retarded null coordinates.

The wave equation for the electromagnetic potential is
\begin{equation}
\left(\boxempty A_{\mu}-\partial_{\mu}\nabla_{\nu}A^{\nu}\right)=0\,.
\end{equation}
Expressing the solution in retarded null coordinates, for simplicity in the gauge $A_{r}=0$, one finds the following set of equations:
\begin{align}
  \partial_{r}^{2}A_{u}+\frac{2}{r}\partial_{r}A_{u}-\partial_{u}\partial_{r}A_{u}+\frac{1}{r^{2}}\left(D^{2}A_{u}-\partial_{u}\left(D^{B}A_{B}\right)\right)&=0\, \label{eq:M1},  \\
  \partial_{r}^{2}A_{u}+\frac{2}{r}\partial_{r}A_{u}-\frac{1}{r^{2}}\partial_{r}\left(D^{B}A_{B}\right)&=0\, ,\label{eq:M2}\\
  \partial_{r}^{2}A_{A}-2\partial_{u}\partial_{r}A_{A}+\partial_{A}\partial_{r}A_{u}+\frac{D^{B}D_{B}A_{A}-D_{A}D_{C}A^{C}-A_{A}}{r^{2}}&=0\,. \label{eq:M3}
\end{align}
Note that, although the gauge $A_r = 0$ has been argued in \cite{Fuentealba:2025ekj} to be improper because it hides the angle-dependent logarithmic gauge transformations \cite{Fuentealba:2022xsz,Fuentealba:2023rvf}, it will be sufficient for our purposes, since we shall focus on the field strength.  Furthermore, even though we work in retarded coordinates, we consider the full vector potential, describing both the retarded and the advanced waves.

The procedure for solving Eqs. \eqref{eq:M1}-\eqref{eq:M3} is described in detail in Appendix \ref{App:Max}. The full expression for the electromagnetic potential (retarded + advanced) then reads:
\begin{align*}
A_{u}  =&-\sum_{l,m}\sum_{k=1}^{l}\frac{1}{2^{k-1}k!r^{k}}\frac{\left(l+k\right)!}{\left(l-k\right)!}\frac{d^{l-k+1}}{du^{l-k+1}}g_{lm}^{\text{ret}}\left(u\right)Y_{l m}\,,\\
 & +\sum_{l,m}\sum_{k=0}^{l}\frac{k}{2^{k}k!r^{k+1}}\frac{\left(l+k\right)!}{\left(l-k\right)!}\left(-\frac{d^{l-k}}{du^{l-k}}g_{lm}^{\text{ret}}\left(u\right)+\left(-1\right)^{k}\frac{d^{l-k}}{dv^{l-k}}g_{lm}^{\text{adv}}\left(v\right)\right)Y_{l m}+\frac{Q}{r}\,.
\end{align*}
\begin{align*}
A_{A}  = &\sum_{l,m}\sum_{k=0}^{l}\frac{1}{2^{k}k!r^{k}}\frac{\left(l+k\right)!}{\left(l-k\right)!}\left(-\frac{d^{l-k}}{du^{l-k}}g_{lm}^{\text{ret}}\left(u\right)+\left(-1\right)^{k}\frac{d^{l-k}}{dv^{l-k}}g_{lm}^{\text{adv}}\left(v\right)\right)\partial_{A}Y_{l m}\\
&+2\sum_{l,m}\frac{d^{l}}{du^{l}}g_{lm}^{\text{ret}}\left(u\right)\partial_{A}Y_{l m}\\
 & +\sum_{l,m}\sum_{k=0}^{l}\frac{1}{2^{k}k!r^{k}}\frac{\left(l+k\right)!}{\left(l-k\right)!}\left(\frac{d^{l-k}}{du^{l-k}}h_{lm}^{\text{ret}}\left(u\right)+\left(-1\right)^{k}\frac{d^{l-k}}{dv^{l-k}}h_{lm}^{\text{adv}}\left(v\right)\right)\sqrt{\gamma}\epsilon_{AB}\partial^{B}Y_{l m}\,.
\end{align*}
The solution depends on the functions $g_{lm}^{\text{ret}}\left(u\right)$ and $g_{lm}^{\text{adv}}\left(v\right)$, associated with the ``longitudinal'' or ``electric'' sector, and $h_{lm}^{\text{ret}}\left(u\right)$ and $h_{lm}^{\text{adv}}\left(v\right)$, associated with the ``transverse'' or ``magnetic'' sector. In the electromagnetic case, there is an additional set of arbitrary functions depending on the retarded and advanced times compared to the scalar field, reflecting the fact that the electromagnetic field carries two local degrees of freedom per spatial point.

From the vector potential, we easily derive the field strength,
\begin{align*}
 F_{ur} =&\sum_{l,m}\sum_{k=0}^{l}\frac{k}{2^{k-1}k!r^{k+1}}\frac{\left(l+k\right)!}{\left(l-k\right)!}\left(-\frac{d^{l-k+1}}{du^{l-k+1}}g_{lm}^{\text{ret}}\left(u\right)-\left(-1\right)^{k}\frac{d^{l-k+1}}{dv^{l-k+1}}g_{lm}^{\text{adv}}\left(v\right)\right)Y_{l m}\\
 & +\sum_{l,m}\sum_{k=0}^{l}\frac{k\left(k+1\right)}{2^{k}k!r^{k+2}}\frac{\left(l+k\right)!}{\left(l-k\right)!}\left(-\frac{d^{l-k}}{du^{l-k}}g_{lm}^{\text{ret}}\left(u\right)+\left(-1\right)^{k}\frac{d^{l-k}}{dv^{l-k}}g_{lm}^{\text{adv}}\left(v\right)\right)Y_{l m}+\frac{Q}{r^{2}}\,,
\end{align*}
\begin{align*}
F_{uA} =&
\sum_{l,m}\sum_{k=0}^{l}
   \frac{1}{2^{k}k!\,r^{k}}
   \frac{(l+k)!}{(l-k)!}
   \Bigg(
      \frac{d^{\,l-k+1}}{du^{\,l-k+1}}g_{lm}^{\text{ret}}(u)
      +(-1)^{k}\frac{d^{\,l-k+1}}{dv^{\,l-k+1}}g_{lm}^{\text{adv}}(v)
   \Bigg)
   \partial_{A}Y_{lm}
\\[4pt]
&+\sum_{l,m}\sum_{k=0}^{l}
   \frac{k}{2^{k}k!\,r^{k+1}}
   \frac{(l+k)!}{(l-k)!}
   \Bigg(
      \frac{d^{\,l-k}}{du^{\,l-k}}g_{lm}^{\text{ret}}(u)
      -(-1)^{k}\frac{d^{\,l-k}}{dv^{\,l-k}}g_{lm}^{\text{adv}}(v)
   \Bigg)
   \partial_{A}Y_{lm}
\\[4pt]
&+\,
      \sum_{l,m}\sum_{k=0}^{l}
         \frac{1}{2^{k}k!\,r^{k}}
         \frac{(l+k)!}{(l-k)!}
         \Big(
            \frac{d^{\,l-k+1}}{du^{\,l-k+1}}h_{lm}^{\text{ret}}(u)
            +(-1)^{k}\frac{d^{\,l-k+1}}{dv^{\,l-k+1}}h_{lm}^{\text{adv}}(v)
         \Big)
        \sqrt{\gamma}\,\epsilon_{AB} \partial^{B}Y_{l m}\,.
\end{align*}

\begin{align*}
F_{rA} =&\sum_{l,m}\sum_{k=0}^{l}\Bigg[
\frac{k}{2^{k}k!\,r^{k+1}}
   \frac{(l+k)!}{(l-k)!}
   \Bigg(
      \frac{d^{l-k}}{du^{l-k}}g_{lm}^{\text{ret}}(u)
      -(-1)^{k}\frac{d^{l-k}}{dv^{l-k}}g_{lm}^{\text{adv}}(v)
   \Bigg) \\
&\quad
\hspace{5.5cm}+\frac{2(-1)^{k}}{2^{k}k!\,r^{k}}
   \frac{(l+k)!}{(l-k)!}
   \frac{d^{l-k+1}}{dv^{l-k+1}}g_{lm}^{\text{adv}}(v)
\Bigg]\partial_{A}Y_{l m} \\[6pt]
&+\sum_{l,m}\sum_{k=0}^{l}\Bigg[
\frac{2(-1)^{k}}{2^{k}k!\,r^{k}}
   \frac{(l+k)!}{(l-k)!}
   \frac{d^{l-k+1}}{dv^{l-k+1}}h_{lm}^{\text{adv}}(v) \\
&\quad
-\frac{k}{2^{k}k!\,r^{k+1}}
   \frac{(l+k)!}{(l-k)!}
   \Bigg(
      \frac{d^{l-k}}{du^{l-k}}h_{lm}^{\text{ret}}(u)
      +(-1)^{k}\frac{d^{l-k}}{dv^{l-k}}h_{lm}^{\text{adv}}(v)
   \Bigg)
\Bigg]\sqrt{\gamma}\,\epsilon_{AB}\partial^{B}Y_{l m}\,.
\end{align*}

\begin{align*}
F_{AB} =&
\sum_{l,m}
   \sum_{k=0}^{l}
      \frac{1}{k!\,(2r)^{k}}
      \frac{(l+k)!}{(l-k)!}
      \Bigg(
         \frac{d^{l-k}}{du^{l-k}}h_{lm}^{\text{ret}}(u)
         +(-1)^{k}\frac{d^{l-k}}{dv^{l-k}}h_{lm}^{\text{adv}}(v)
      \Bigg)
 \nonumber\\
&\times
\Big(
\epsilon_{BC}\,\partial_{A}\partial^{C}Y_{l m}
   -\epsilon_{AC}\,\partial_{B}\partial^{C}Y_{l m}
\Big)\,.
\end{align*}

Once the electromagnetic potential is known in retarded coordinates, one can derive its form in advanced coordinates by making the corresponding change of coordinates.  However, one must also perform a change of gauge if one wants the resulting potential to fulfill the gauge conditions $A_r=0$, $A_v = O(1/r)$.  Such a complication does not occur for the gauge-invariant field strength, which we shall thus simply derive in advanced coordinates from their expression in retarded coordinates through the corresponding change of variables.

\subsection{Finite energy flux conditions}

As in the scalar case, the functions of $u$ and $v$ appearing in the general solution are not arbitrary but constrained by finite energy flux conditions.

The energy flux is determined by the integral of the Pointing vector
\[
\frac{dE}{dt}=\int_{s^{2}}ds_{j}\,E^{i}F_{ij}\,.
\]
Let us write the above expression in spherical coordinates: 
\[
\frac{dE}{dt}=\int_{s^{2}}d^{2}\hat{x}\,\sqrt{\gamma}F_{Ar}E^{A}\,.
\]
where 
\begin{align*}
F_{Ar} & =\partial_{A}A_{r}-\partial_{r}A_{A}\,,\\
E^{A} & =\gamma^{AB}\left(\partial_{A}A_{t}-\partial_{t}A_{A}\right) \, .
\end{align*}
Here, the components are evaluated in $(t, r, x^A)$ coordinates. For the retarded component of the solution, one has \footnote{Using $A_{r}=0$ and $A_{u}=O\left(r^{-1}\right)$ in $(u, r, x^A)$ coordinates implies $ A_r = O\left(r^{-1}\right)$,  $A_t = O\left(r^{-1}\right)$ in $(t, r, x^A)$ coordinates}
\begin{align*}
F_{Ar} & =\partial_{A}A_{r}\left(t-r,\hat{x}\right)-\partial_{r}A_{A}\left(t-r,\hat{x}\right)=\partial_{u}\bar{A}_{A}\left(u,\hat{x}\right)+O\left(r^{-1}\right)\,,\\
E^{A} & =\gamma^{AB}\left(\partial_{A}A_{t}-\partial_{t}A_{A}\right)=-\gamma^{AB}\partial_{u}\bar{A}_{A}\left(u,\hat{x}\right)+O\left(r^{-1}\right),
\end{align*}
where 
\[
A_{A}\left(u,r,\hat{x}\right)=\bar{A}_{A}\left(u,\hat{x}\right)+O\left(r^{-1}\right)
\]
One then gets
\[
\frac{dE}{du}=-\int_{s^{2}}d^{2}\hat{x}\,\sqrt{\gamma}\gamma^{AB}\left(\partial_{u}\bar{A}_{A}\right)\left(\partial_{u}\bar{A}_{B}\right)\,.
\]
Therefore, the total radiated energy through $\mathscr{I}^{+}$ by the retarded component of the electromagnetic field is
given by
\[
\left.\Delta E\right|_{\mathscr{I}^{+}}=-\int_{\mathscr{I}^{+}}dud^{2}\hat{x}\,\sqrt{\gamma}\gamma^{AB}\left(\partial_{u}\bar{A}_{A}\right)\left(\partial_{u}\bar{A}_{B}\right)\,.
\]
Requiring this expression to be finite imposes, as in the scalar field case,
\begin{align}
\lim_{u\rightarrow\pm\infty}\frac{d^{l}}{du^{l}}g_{lm}^{\text{ret}}\left(u\right) & =g_{lm}^{\pm\text{ret}}\log\left(\pm u\right)+G_{lm}^{\pm\text{ret}}+\dots\,,\label{eq:Fin_ret_1}\\
\lim_{u\rightarrow\pm\infty}\frac{d^{l}}{du^{l}}h_{lm}^{\text{ret}}\left(u\right) & =h_{lm}^{\pm\text{ret}}\log\left(\pm u\right)+H_{lm}^{\pm\text{ret}}+\dots\,,\label{eq:Fin_ret_2}
\end{align}
where $g_{lm}^{\text{ret}}$, $G_{lm}^{\pm\text{ret}}$, $h_{lm}^{\text{ret}}$ and $H_{lm}^{\pm\text{ret}}$ are constants.  It follows by integration that
\begin{align*}
\frac{d^{l-k}}{du^{l-k}}g_{lm}^{\text{ret}}\left(u\right) & \underset{u\rightarrow-\infty}{=}g_{lm}^{-\text{ret}}\left(\frac{u^{k}}{k!}\log\left(-u\right)-\frac{1}{k!}H_{k}u^{k}\right)+G_{lm}^{-\text{ret}}\frac{u^{k}}{k!}+\dots\,,\\
\frac{d^{l-k}}{du^{l-k}}h_{lm}^{\text{ret}}\left(u\right) & \underset{u\rightarrow-\infty}{=}h_{lm}^{-\text{ret}}\left(\frac{u^{k}}{k!}\log\left(-u\right)-\frac{1}{k!}H_{k}u^{k}\right)+H_{lm}^{-\text{ret}}\frac{u^{k}}{k!}+\dots\,.
\end{align*}
Here, $H_{k}=\sum_{n=1}^{k}\frac{1}{n}$
represents the harmonic numbers. The integration constants have been omitted, as they contribute only subleading terms and are therefore irrelevant for the present analysis (see formulas in appendix {\bf \ref{App:Gen}}).

Analogously, finiteness of the total incoming energy through $\mathscr{I}^{-}$ from the advanced field yields the conditions
\begin{align}
\lim_{v\rightarrow\pm\infty}\frac{d^{l}}{dv^{l}}g_{lm}^{\text{adv}}\left(v\right) & =g_{lm}^{\pm\text{adv}}\log\left(\pm v\right)+G_{lm}^{\pm\text{adv}}+\dots\,,\label{eq:Fin_adv_1}\\
\lim_{v\rightarrow\pm\infty}\frac{d^{l}}{dv^{l}}h_{lm}^{\text{adv}}\left(v\right) & =h_{lm}^{\pm\text{adv}}\log\left(\pm v\right)+H_{lm}^{\pm\text{adv}}+\dots\,,\label{eq:Fin_adv_2}
\end{align}
from which one gets
\begin{align}
\frac{d^{l-k}}{dv^{l-k}}g_{lm}^{\text{adv}}\left(v\right) & \underset{v\rightarrow+\infty}{=}g_{lm}^{+\text{adv}}\left(\frac{v^{k}}{k!}\log\left(v\right)-\frac{1}{k!}H_{k}v^{k}\right)+G_{lm}^{+\text{adv}}\frac{v^{k}}{k!}+\dots\,,\label{eq:Int_Fin_adv_1}\\
\frac{d^{l-k}}{dv^{l-k}}h_{lm}^{\text{adv}}\left(v\right) & \underset{v\rightarrow+\infty}{=}h_{lm}^{+\text{adv}}\left(\frac{v^{k}}{k!}\log\left(v\right)-\frac{1}{k!}H_{k}v^{k}\right)+H_{lm}^{+\text{adv}}\frac{v^{k}}{k!}+\dots\,\label{eq:Int_Fin_adv_2}.
\end{align}

\subsection{Asymptotic behaviour near $i^0$}
Further conditions arise from the requirement that the field strength involves no logarithm at spatial infinity.  These relate the advanced functions to the retarded functions and will be key in the derivation of the matching conditions below.

We shall work out the components $\tilde{F}_{\mu \nu}$ of the electromagnetic tensor in $\left(t,r,x^{A}\right)$ coordinates from the components in retarded coordinates. One has
\[
\tilde{F}_{tr}=F_{ur}\,,\qquad\qquad\tilde{F}_{tA}=F_{uA}\,,\qquad\qquad\tilde{F}_{rA}=F_{rA}-F_{uA}\,,\qquad\qquad\tilde{F}_{AB}=F_{AB}\,.
\]
Since our goal is to determine the conditions that enforce the absence of $\log r$ terms at spatial infinity, we trace only the $\log u$ and $\log v$ terms in the field strength, since these are the only source of $\log r$ terms.  

Let us first focus on the component $\tilde{F}_{tr}= F_{ur}$.  Using the relations $u=t-r$ and $v=t+r$ as well as the expressions for the retarded and advanced functions resulting from the finite radiated/incoming energy conditions, we find (keeping only the logs)
\begin{multline*}
\tilde{F}_{tr}  =\sum_{l,m}\sum_{k=0}^{l}\frac{k}{2^{k-1}k!r^{k+1}}\frac{\left(l+k\right)!}{\left(l-k\right)!}\left(-g_{lm}^{-\text{ret}}\frac{k\left(t-r\right)^{k-1}}{k!}\log\left(r-t\right)-\right. \\  \left.\left(-1\right)^{k}g_{lm}^{+\text{adv}}k\frac{\left(t+r\right)^{k-1}}{k!}\log\left(t+r\right)\right)Y_{lm}\\
+\sum_{l,m}\sum_{k=0}^{l}\frac{k\left(k+1\right)}{2^{k}k!r^{k+2}}\frac{\left(l+k\right)!}{\left(l-k\right)!}\left(\left(-1\right)^{k}g_{lm}^{+\text{adv}}\frac{\left(t+r\right)^{k}}{k!}\log\left(t+r\right)-g_{lm}^{-\text{ret}}\frac{\left(t-r\right)^{k}}{k!}\log\left(r-t\right)\right)Y_{lm}.
\end{multline*}
Therefore, in the limit when $r\rightarrow\infty$, with $t$ held
fixed, we find
\[
\left.\tilde{F}_{tr}\right|_{i^{0}}=\frac{\log r}{r^{2}}\sum_{l,m} \left(\sum_{k=0}^{l}\frac{\left(-1\right)^{k}k\left(k-1\right)}{2^{k}\left(k!\right)^{2}}\frac{\left(l+k\right)!}{\left(l-k\right)!}\right)\left(g_{lm}^{-\text{ret}}-g_{lm}^{+\text{adv}}\right)Y_{lm}+\cdots\,.
\]
Expressing the sum in terms of the derivatives of the Legendre polynomials $P_l(x)$ evaluated at zero (see \eqref{Pl20}) we can write
\begin{align}
\sum_{k=0}^{l}\frac{\left(-1\right)^{k}k\left(k-1\right)}{2^{k}\left(k!\right)^{2}}\frac{\left(l+k\right)!}{\left(l-k\right)!}=P''_l(0).\label{Pprimaprima0}
\end{align}
Notice that the above relation vanishes for odd values of $l$, and is non-zero for even values
of $l$. 
Let us now compute $\tilde{F}_{tA} = F_{uA}$.
Following the same rules as for $\tilde{F}_{tr}$, we find

\begin{align*}
\tilde{F}_{tA} & =\sum_{l,m}\sum_{k=0}^{l}\frac{1}{2^{k}k!r^{k}}\frac{\left(l+k\right)!}{\left(l-k\right)!}\left(g_{lm}^{-\text{ret}}\frac{k\left(t-r\right)^{k-1}}{k!}\log\left(r-t\right)\right. \\
&\;\;\;\;\;\;\;\;\;\;\;\;\;\;\;\;\;\;\;\;\;\;\;\;\;\;\;\;\;\;\;\;\;\;\;\;\;\;\;\;\;\;\;\;\;\;\;\;\;\;\;\;\;\;\;\;\;\;\;\;\;+\left.\left(-1\right)^{k}g_{lm}^{+\text{adv}}k\frac{\left(t+r\right)^{k-1}}{k!}\log\left(t+r\right)\right)\partial_{A}Y_{lm}\\
 & +\sum_{l,m}\sum_{k=0}^{l}\frac{k}{2^{k}k!r^{k+1}}\frac{\left(l+k\right)!}{\left(l-k\right)!}\left(g_{lm}^{-\text{ret}}\frac{\left(t-r\right)^{k}}{k!}\log\left(r-t\right)\right. \\
&\left.\;\;\;\;\;\;\;\;\;\;\;\;\;\;\;\;\;\;\;\;\;\;\;\;\;\;\;\;\;\;\;\;\;\;\;\;\;\;\;\;\;\;\;\;\;\;\;\;\;\;\;\;\;\;\;\;\;\;\;\;\;\; -\left(-1\right)^{k}g_{lm}^{+\text{adv}}\frac{\left(t+r\right)^{k}}{k!}\log\left(t+r\right)\right)\partial_{A}Y_{lm}\\
&+\sum_{l,m}\sum_{k=0}^{l}\frac{1}{2^{k}k!r^{k}}\frac{\left(l+k\right)!}{\left(l-k\right)!}\left(h_{lm}^{-\text{ret}}\frac{\left(t-r\right)^{k-1}}{k!}\log\left(r-t\right)\right.\\
&\;\;\;\;\;\;\;\;\;\;\;\;\;\;\;\;\;\;\;\;\;\;\;\;\;\;\;\;\;\;\;\;\;\;\;\;\;\;\;\;\;\;\;\;\;\;\;\;\;\;\;\left.  +\left(-1\right)^{k}h_{lm}^{+\text{adv}}k\frac{\left(t+r\right)^{k-1}}{k!}\log\left(t+r\right)\right)\sqrt{\gamma}\epsilon_{AB}\partial^{B}Y_{lm}.
\end{align*}
Therefore, taking the limit when $r\rightarrow\infty$, keeping $t$
fixed, one finds
\begin{align*}
\tilde{F}_{tA} & =\frac{\log r}{r}\left[\sum_{l,m}\sum_{k=0}^{l}\frac{\left(-1\right)^{k}k}{2^{k}\left(k!\right)^{2}}\frac{\left(l+k\right)!}{\left(l-k\right)!}\left(h_{lm}^{+\text{adv}}-h_{lm}^{-\text{ret}}\right)\sqrt{\gamma}\epsilon_{AB}\partial^{B} Y_{lm}\right]+\dots\,.
\end{align*}
Expressing the sum in terms of the derivatives of the Legendre polynomials evaluated at zero (see equation \eqref{Pl10} for a derivation),
\begin{equation}
\label{Pprima0}
\sum_{k=0}^{l}\frac{\left(-1\right)^{k}k}{2^{k}\left(k!\right)^{2}}\frac{\left(l+k\right)!}{\left(l-k\right)!}= -2P'_{l}(0)\,, 
\end{equation}
which now vanishes for even values of $l$, we obtain
\[
\left.\tilde{F}_{tA}\right|_{i^{0}}=\frac{\log r}{r}\left[-2\sum_{l,m}P'_{l}(0)\left(h_{lm}^{+\text{adv}}-h_{lm}^{-\text{ret}}\right)\sqrt{\gamma}\epsilon_{AB}\partial^{B}Y_{lm}\right]+\dots\,.
\]
Similar considerations yield $\tilde{F}_{rA}=F_{rA}-F_{uA}$,
\begin{align*}
\tilde{F}_{rA} =
\sum_{l,m}
\Bigg[
   \sum_{k=0}^{l}
      \frac{1}{2^{k}k!\,r^{k}}
      \frac{(l+k)!}{(l-k)!}
      \Bigg(
         -g_{lm}^{-\text{ret}}
         \frac{k\,(t-r)^{k-1}}{k!}\log(r-t)
         \\&\hspace{-4cm}+(-1)^{k}g_{lm}^{+\text{adv}}
         \frac{k\,(t+r)^{k-1}}{k!}\log(t+r)
      \Bigg)
\Bigg]\partial_{A}Y_{lm}
\\
+\sum_{l,m}\sum_{k=0}^{l}
   \frac{1}{2^{k}k!\,r^{k}}
   \frac{(l+k)!}{(l-k)!}
   \Bigg(
      -h_{lm}^{-\text{ret}}
      \frac{k\,(t-r)^{k-1}}{k!}\log(r-t)
      \\&\hspace{-4cm}+(-1)^{k}h_{lm}^{+\text{adv}}
      \frac{k\,(t+r)^{k-1}}{k!}\log(t+r)
   \Bigg)
\sqrt{\gamma}\,\epsilon_{AB}\,\partial^{B}Y_{lm}
\\
-\sum_{l,m}
\Bigg[
   \sum_{k=0}^{l}
      \frac{k}{2^{k}k!\,r^{k+1}}
      \frac{(l+k)!}{(l-k)!}
      \Bigg(
         h_{lm}^{-\text{ret}}
         \frac{(t-r)^{k}}{k!}\log(r-t)
       \\&\hspace{-4cm}+(-1)^{k}h_{lm}^{+\text{adv}}
         \frac{(t+r)^{k}}{k!}\log(t+r)
      \Bigg)
\Bigg]
\sqrt{\gamma}\,\epsilon_{AB}\,\partial^{B}Y_{lm}\,.
\end{align*}
After using \eqref{Pprima0}, and taking the limit $r\rightarrow\infty$ with $t$ kept constant, reduces to
\begin{align*}
\left.\tilde{F}_{rA}\right|_{i^0} & =\frac{\log r}{r}\sum_{l,m}\left[-2P_l'(0)\left(g_{lm}^{-\text{ret}}+g_{lm}^{+\text{adv}}\right)\right]\partial_{A}Y_{lm}+\dots\,.
\end{align*}
Finally, let us turn to  $\tilde{F}_{AB} = F_{AB}$.
We find this time by performing similar manipulations
\begin{multline*}
F_{AB}=\sum_{l,m}\sum_{k=0}^{l}\frac{\left(l+k\right)!}{\left(l-k\right)!}\left(h_{lm}^{-\text{ret}}\frac{\left(t-r\right)^{k}}{k!^2\left(2r\right)^{k}}\log\left(r-t\right)+h_{lm}^{+\text{adv}}\frac{\left(-1\right)^{k}\left(t+r\right)^{k}}{k!^2\left(2r\right)^{k}}\log\left(t+r\right)\right)\\\times\left(\epsilon_{BC}\partial_{A}\partial^{C}Y_{lm}-\epsilon_{AC}\partial_{B}\partial^{C}Y_{lm}\right).
\end{multline*}
Therefore, in the limit when $r\rightarrow\infty$ keeping $t$ constant,
we obtain
\[
F_{AB}=\log r\sum_{l,m}\left[P_l(0)\left(h_{lm}^{-\text{ret}}+h_{lm}^{+\text{adv}}\right)\right]\left(\epsilon_{BC}\partial_{A}\partial^{C}Y_{lm}-\epsilon_{AC}\partial_{B}\partial^{C}Y_{lm}\right), 
\]
where we used the relation \eqref{Pl0} 
\begin{align}
    \sum_{k=0}^l \frac{(-1)^k}{(k!)^2}\frac{(l+k)!}{(l-k)!}=P_l(0). \label{P0} 
\end{align}

In order to eliminate the leading $\log r$ terms, we must impose the following requirements
\begin{equation}
P_l'(0)\left(h_{lm}^{+\text{adv}}-h_{lm}^{-\text{ret}}\right) =0,\quad
P_l(0)\left(h_{lm}^{-\text{ret}}+h_{lm}^{+\text{adv}}\right)  =0,
\end{equation}
and 
\begin{equation}
P'_l(0)\left(g_{lm}^{-\text{ret}}+g_{lm}^{+\text{adv}}\right)=0,
\quad
P''_l(0)\left(g_{lm}^{-\text{ret}}-g_{lm}^{+\text{adv}}\right)=0.
\end{equation}
These have the same form as in the scalar case discussed in \eqref{eq:cond1} and \eqref{eq:cond2}. They lead to the following conditions,
\begin{equation}
g_{lm}^{-\text{ret}}=\left(-1\right)^{l}g_{lm}^{+\text{adv}}\,,\qquad\qquad h_{lm}^{-\text{ret}}=-\left(-1\right)^{l}h_{lm}^{+\text{adv}}\,.\label{eq:match}
\end{equation}
These relations between the advanced and retarded fields are essential for establishing the matching conditions connecting past and future null infinity for the logarithmic terms.

Note that, as in the scalar field case, the contributions to the leading logarithmic term at $i^{0}$ arise from all powers of $r$ appearing in the electromagnetic tensor expressed in retarded or advanced coordinates. Accordingly, it becomes necessary to extend the near-$\mathcal{I^{\pm}}$ expansion into the bulk in order to derive the conditions \eqref{eq:match} and determine the leading orders in the vicinity of $i^{0}$.

\subsection{Matching conditions}
The matching conditions of \cite{He:2014cra,Kapec:2015ena,Strominger:2017zoo} relate the leading order of the radial electric and magnetic fields at the past of $\mathscr{I}^{+}$ to the leading order of the corresponding fields at the future of $\mathscr{I}^{-}$. They were derived assuming no logarithmic terms at null infinity.  To derive them in the general case, let us thus examine the behaviour of the fields near null infinity, starting with the electromagnetic potential.

\subsubsection{Expansion of the electromagnetic potential near $\mathscr{I}^{+}$}

The advanced solution is by definition the $v$-dependent part of the electromagnetic potential, i.e.,
\begin{align*}
A_{u}^{\text{adv}}  =&\sum_{l,m}\sum_{k=0}^{l}\frac{\left(-1\right)^{k}k}{2^{k}k!r^{k+1}}\frac{\left(l+k\right)!}{\left(l-k\right)!}\frac{d^{l-k}}{dv^{l-k}}g_{lm}^{\text{adv}}\left(v\right)Y_{lm}\,,\\
A_{A}^{\text{adv}}  =&\sum_{l,m}\sum_{k=0}^{l}\frac{\left(-1\right)^{k}}{2^{k}k!r^{k}}\frac{\left(l+k\right)!}{\left(l-k\right)!}\left(\frac{d^{l-k}}{dv^{l-k}}g_{lm}^{\text{adv}}\left(v\right)\right)\partial_{A}Y_{lm}\\
&+\sum_{l,l}\sum_{k=0}^{l}\frac{\left(-1\right)^{k}}{2^{k}k!r^{k}}\frac{\left(l+k\right)!}{\left(l-k\right)!}\left(\frac{d^{l-k}}{dv^{l-k}}h_{lm}^{\text{adv}}\left(v\right)\right)\sqrt{\gamma}\epsilon_{AB}\partial^{B}Y_{lm}\,.
\end{align*}
When expanded near future null infinity, it brings $\log r$ terms, as in the scalar case, which we now determine.

Using the expressions \eqref{eq:Fin_adv_1}-\eqref{eq:Int_Fin_adv_2} for $g_{lm}^{\text{adv}}$ and $h_{lm}^{\text{adv}}$, the advanced solution becomes 
\begin{align*}
A_{u}^{\text{adv}} & =\sum_{l,m}\sum_{k=0}^{l}\frac{\left(-1\right)^{k}k}{2^{k}k!^2r^{k+1}}\frac{\left(l+k\right)!}{\left(l-k\right)!}\left[g_{lm}^{+\text{adv}}\left(v^{k}\log\left(v\right)-H_{k}v^{k}\right)+G_{lm}^{+\text{adv}}v^{k}+\dots\right]Y_{lm}\,,\\
A_{A}^{\text{adv}} & =\sum_{l,m}\sum_{k=0}^{l}\frac{\left(-1\right)^{k}}{2^{k}k!^2r^{k}}\frac{\left(l+k\right)!}{\left(l-k\right)!}\left[g_{lm}^{+\text{adv}}\left(v^{k}\log\left(v\right)-H_{k}v^{k}\right)+G_{lm}^{+\text{adv}}v^{k}+\dots\right]\partial_{A}Y_{lm}\\
 & +\sum_{l,m}\sum_{k=0}^{l}\frac{\left(-1\right)^{k}}{2^{k}k!^2r^{k}}\frac{\left(l+k\right)!}{\left(l-k\right)!}\left[h_{lm}^{+\text{adv}}\left(v^{k}\log\left(v\right)-H_{k}v^{k}\right)+H_{lm}^{+\text{adv}}v^{k}+\dots\right]\sqrt{\gamma}\epsilon_{AB}\partial^{B}Y_{lm}.
\end{align*}
Substituting $v=u+2r$, and expanding for large values of $r$,
i.e., near future null infinity, we find
\begin{align*}
A_{u}^{\text{adv}} & =\frac{\log r}{r}\left[\sum_{l,m}g_{lm}^{+\text{adv}}\left(\sum_{k=0}^{l}\frac{\left(-1\right)^{k}k}{\left(k!\right)^{2}}\frac{\left(l+k\right)!}{\left(l-k\right)!}\right)Y_{lm}\right]+\\
&\,\,\,\,\,\,\,\,\,\,\,\frac{1}{r}\sum_{l,m}\sum_{k=0}^{l}\frac{\left(-1\right)^{k}k}{\left(k!\right)^{2}}\frac{\left(l+k\right)!}{\left(l-k\right)!}\left[g_{lm}^{+\text{adv}}\log2-g_{lm}^{+\text{adv}}H_{k}+G_{lm}^{+\text{adv}}\right]Y_{lm}+\dots\,,\\
A_{A}^{\text{adv}} & =\log r\, \sum_{l,m} \left(\sum_{k=0}^{l}\frac{\left(-1\right)^{k}}{\left(k!\right)^{2}}\frac{\left(l+k\right)!}{\left(l-k\right)!}\right) \left[g_{lm}^{+\text{adv}}\partial_{A}Y_{lm}+h_{lm}^{+\text{adv}}\sqrt{\gamma}\epsilon_{AB}\partial^{B}Y_{lm}\right]\\
&+\sum_{l,m}\sum_{k=0}^{l}\frac{\left(-1\right)^{k}}{\left(k!\right)^{2}}\frac{\left(l+k\right)!}{\left(l-k\right)!}\left[g_{lm}^{+\text{adv}}\log2-g_{lm}^{+\text{adv}}H_{k}+G_{lm}^{+\text{adv}}\right]\partial_{A}Y_{lm} \\
&+\sum_{l,m}\sum_{k=0}^{l}\frac{\left(-1\right)^{k}}{\left(k!\right)^{2}}\frac{\left(l+k\right)!}{\left(l-k\right)!}\left[h_{lm}^{+\text{adv}}\log2-h_{lm}^{+\text{adv}}H_{k}+H_{lm}^{+\text{adv}}\right]\sqrt{\gamma}\epsilon_{AB}\partial^{B}Y_{lm}\\
 & +\frac{\log r}{r}\left[\sum_{l,m}\sum_{k=0}^{l}\frac{\left(-1\right)^{k}k}{2\left(k!\right)^{2}}\frac{\left(l+k\right)!}{\left(l-k\right)!}u\left(g_{lm}^{+\text{adv}}\partial_{A}Y_{lm}+ h_{lm}^{+\text{adv}}\sqrt{\gamma}\epsilon_{AB}\partial^{B}Y_{lm}\right)\right]+\dots\,.
\end{align*}
The above expressions can be simplified using the identities \eqref{Plm1} and \eqref{Plm2}
\begin{align}
\sum_{k=0}^{l}\frac{\left(-1\right)^{k}}{\left(k!\right)^{2}}\frac{\left(l+k\right)!}{\left(l-k\right)!}=\left(-1\right)^{l}\,,\qquad\sum_{k=0}^{l}\frac{\left(-1\right)^{k}k}{\left(k!\right)^{2}}\frac{\left(l+k\right)!}{\left(l-k\right)!}=\left(-1\right)^{l}l\left(l+1\right)\,,\label{eq:Id}
\end{align}
which imply
\begin{align*}
\left.A_{u}^{\text{adv}}\right|_{\mathscr{I}^{+}} & =\frac{\log r}{r}\left[\sum_{l,m}\left(-1\right)^{l}l\left(l+1\right)g_{lm}^{+\text{adv}}Y_{lm}\right]+\\
&\frac{1}{r}\sum_{l,m}\left[\left(-1\right)^{l}l\left(l+1\right)\left(g_{lm}^{+\text{adv}}\log2+G_{lm}^{+\text{adv}}\right)-\sum_{k=0}^{l}g_{lm}^{+\text{adv}}\frac{\left(-1\right)^{k}k}{\left(k!\right)^{2}}\frac{\left(l+k\right)!}{\left(l-k\right)!}H_{k}\right]Y_{lm}\,,\\
\left.A_{A}^{\text{adv}}\right|_{\mathscr{I}^{+}} & =\log r\sum_{l,m} \left[\left(-1\right)^{l}g_{lm}^{+\text{adv}}\partial_{A}Y_{lm}+\left(-1\right)^{l}h_{lm}^{+\text{adv}}\sqrt{\gamma}\epsilon_{AB}\partial^{B}Y_{lm}\right]\\
 & +\sum_{l,m}\left[\left(-1\right)^{l}\left(g_{lm}^{+\text{adv}}\log2+G_{lm}^{+\text{adv}}\right)-g_{lm}^{+\text{adv}}\sum_{k=0}^{l}\frac{\left(-1\right)^{k}}{\left(k!\right)^{2}}\frac{\left(l+k\right)!}{\left(l-k\right)!}H_{k}\right]\partial_{A}Y_{lm}+\\
 & +\sum_{l,m}\left[\left(-1\right)^{l}\left(h_{lm}^{+\text{adv}}\log2+H_{lm}^{+\text{adv}}\right)-h_{lm}^{+\text{adv}}\sum_{k=0}^{l}\frac{\left(-1\right)^{k}}{\left(k!\right)^{2}}\frac{\left(l+k\right)!}{\left(l-k\right)!}H_{k}\right]\sqrt{\gamma}\epsilon_{AB}\partial^{B}Y_{lm}\\
 & +\frac{\log r}{r}\left[\sum_{l,m}\frac{u}{2}\left(-1\right)^{l}l\left(l+1\right)\left(g_{lm}^{+\text{adv}}\partial_{A}Y_{lm}+h_{lm}^{+\text{adv}}\sqrt{\gamma}\epsilon_{AB}\partial^{B}Y_{lm}\right) \right]+\dots\,.
\end{align*}

Therefore, the full solution (retarded + advanced) near $\mathscr{I}^{+}$ is given by
\begin{align}
A_{u} & =\frac{\log r}{r}\left[\sum_{l,m}\left(-1\right)^{l}l\left(l+1\right)g_{lm}^{+\text{adv}}Y_{lm}\right]+\nonumber\\
&\frac{1}{r}\sum_{l,m}\left[\left(-1\right)^{l}l\left(l+1\right)\left(g_{lm}^{+\text{adv}}\log2+G_{lm}^{+\text{adv}}\right)-g_{lm}^{+\text{adv}}\sum_{k=0}^{l}\left(\frac{\left(-1\right)^{k}k}{\left(k!\right)^{2}}\frac{\left(l+k\right)!}{\left(l-k\right)!}H_{k}\right)\right] \label{eq:AU}\\
 & \left.-\sum_{l,m}\sum_{k=1}^{l}l\left(l+1\right)\left(\frac{d^{l}}{du^{l}}g_{lm}^{\text{ret}}\left(u\right)\right)+Q\right]Y_{lm}+\dots\,,\nonumber
\end{align}
and
\begin{align}
A_{A} &= \log r\Bigg[
  \sum_{l,m} (-1)^{l} g_{lm}^{+\text{adv}}\,\partial_{A}Y_{lm}
  + \sum_{l,m} (-1)^{l} h_{lm}^{+\text{adv}}\,\sqrt{\gamma}\,\epsilon_{AB}\,\partial^{B}Y_{lm}
\Bigg] \nonumber\\
& \quad
+ \sum_{l,m}\Bigg[
  \frac{d^{l}}{du^{l}}\,g_{lm}^{\text{ret}}(u)
  + (-1)^{l}\!\left(g_{lm}^{+\text{adv}}\log 2 + G_{lm}^{+\text{adv}}\right) 
  \nonumber\\ 
  & \,\,\,\,\,\,\,\,\,\,\,\,\,\,\,\,\,\,\,\,\,\,\,\,\,\,\,\,\,\,\,\,\,\,\,\,\,\,\,\,\,\,\,\,\,\,\,\,\,\,\,\,\,\,\,\,\,\,\,\,\,\,\,\,\,\,\,\,\,\,\,\,\,\,\,\,\,\,\,\,\,\,\,\,\,\,\,\,\,\,\,\,\,\,\,\,\,\,\,\,\,- g_{lm}^{+\text{adv}}\!\sum_{k=0}^{l}\frac{(-1)^{k}}{(k!)^{2}}\frac{(l+k)!}{(l-k)!}\,H_{k}
\Bigg]\partial_{A}Y_{lm} \label{eq:AR}\\
&\quad
+ \sum_{l,m}\Bigg[
  \frac{d^{l}}{du^{l}}\,h_{lm}^{\text{ret}}(u)
  + (-1)^{l}\!\left(h_{lm}^{+\text{adv}}\log 2 + H_{lm}^{+\text{adv}}\right)
\nonumber \\
& \,\,\,\,\,\,\,\,\,\,\,\,\,\,\,\,\,\,\,\,\,\,\,\,\,\,\,\,\,\,\,\,\,\,\,\,\,\,\,\,\,\,\,\,\,\,\,\,\,\,\,\,\,\,\,\,\,\,\,\,\,\,\,\,\,\,\,\,\,\,
- h_{lm}^{+\text{adv}}\!\sum_{k=0}^{l}\frac{(-1)^{k}}{(k!)^{2}}\frac{(l+k)!}{(l-k)!}\,H_{k}
\Bigg]\sqrt{\gamma}\,\epsilon_{AB}\,\partial^{B}Y_{lm}\nonumber\\
&\quad
+ \frac{\log r}{r}\Bigg[
  \sum_{l,m}\frac{u}{2}(-1)^{l}l(l+1)
  \big( g_{lm}^{+\text{adv}}\,\partial_{A}Y_{lm}
  + h_{lm}^{+\text{adv}}\,\sqrt{\gamma}\,\epsilon_{AB}\,\partial^{B}Y_{lm} \big)
\Bigg] + \dots \,.\nonumber
\end{align}

\subsubsection{Electromagnetic tensor near $\mathscr{I}^{+}$}\label{subsec:Electromagnetic-tensor}
Let us compute the electromagnetic tensor near $\mathscr{I}^{+}$, focusing on the radial electric field $F_{ur}$ and the radial magnetic field $F_{AB}$ occurring in the matching conditions of \cite{He:2014cra,Kapec:2015ena,Strominger:2017zoo}.  These components are scalars under the changes of coordinates $(u,r) \leftrightarrow (t,r) \leftrightarrow (v,r)$.  We can also derive the matching conditions for the other components since we have all the expressions, but because there is nothing we can compare them with in the literature, we shall not do so here. 
 
From the asymptotic developments of the vector potential in Eqs. \eqref{eq:AU} and \eqref{eq:AR}, one gets
\begin{align}
F_{ur} & =\frac{\log r}{r^{2}}\left[\sum_{l,m}\left(-1\right)^{l}l\left(l+1\right)g_{lm}^{+\text{adv}}Y_{lm}\right]\nonumber\\
\hspace{-1cm}&+\frac{1}{r^{2}}\sum_{l,m}\left[\left(-1\right)^{l}l\left(l+1\right)(g_{lm}^{+\text{adv}}\left(\log2-1\right)+G_{lm}^{+\text{adv}})-\sum_{k=0}^{l}g_{lm}^{+\text{adv}}\frac{\left(-1\right)^{k}H_{k} k\left(l+k\right)!}{\left(k!\right)^{2}\left(l-k\right)!}\right.\nonumber \\
 & \left.-\sum_{k=1}^{l}l\left(l+1\right)\left(\frac{d^{l}}{du^{l}}g_{lm}^{\text{ret}}\left(u\right)\right)\right]Y_{lm}+ \frac{Q}{r^2}+\dots\,,\label{eq:Furlog}
\end{align}
and
\begin{align*}
F_{AB} & =\log r\sum_{l,m}\left(-1\right)^{l}h_{lm}^{+\text{adv}}\sqrt{\gamma}\left(\epsilon_{BC}\partial_{A}\partial^{C}Y_{lm}-\epsilon_{AC}\partial_{B}\partial^{C}Y_{lm}\right)\\
 & +\sum_{l,m}\left[\frac{d^{l}}{du^{l}}h_{lm}^{\text{ret}}\left(u\right)+\left(-1\right)^{l}\left(h_{lm}^{+\text{adv}}\log2+H_{lm}^{+\text{adv}}\right)-h_{lm}^{+\text{adv}}\left(\sum_{k=0}^{l}\frac{\left(-1\right)^{k}}{\left(k!\right)^{2}}\frac{\left(l+k\right)!}{\left(l-k\right)!}H_{k}\right)\right]\\
&\times\sqrt{\gamma}\left(\epsilon_{BC}\partial_{A}\partial^{C}Y_{lm}-\epsilon_{AC}\partial_{B}\partial^{C}Y_{lm}\right)+\dots\,.
\end{align*}

In both components of the field strength, there is a leading radial logarithmic contribution arising from the advanced wave. In particular, the leading term of $F_{ur}$ is entirely determined by the ``electric'' mode $g_{lm}^{+\text{adv}}$ of the advanced wave, while the leading term of $F_{AB}$ depends only on the ``magnetic'' advanced mode $h_{lm}^{+\text{adv}}$ of the electromagnetic wave.

\subsubsection{Electromagnetic tensor  near $\mathscr{I}^{-}$}
Similar developments, in which one inserts $u= v- 2r$ into the retarded solution, yield the asymptotic expansion of the field strength near  $\mathscr{I}^{-}$,
\begin{align}
F_{vr} & =\frac{\log r}{r^{2}}\sum_{l,m}\left(-1\right)^{l+1}l\left(l+1\right)g_{lm}^{-\text{ret}}Y_{lm}\nonumber \\
 & -\frac{1}{r^{2}}\sum_{l,m}\left[-l\left(l+1\right)\frac{d^{l}}{dv^{l}}g_{lm}^{\text{adv}}\left(v\right)Y_{lm}+\left(-1\right)^{l}l\left(l+1\right)\left(g_{lm}^{-\text{ret}}\left(\log2-1\right)+G_{lm}^{-\text{ret}}\right)\right. \label{eq:Fvrlog}\\
 &\left.-\left(\sum_{k=0}^{l}\frac{\left(-1\right)^{k}k}{\left(k!\right)^{2}}\frac{\left(l+k\right)!}{\left(l-k\right)!}H_{k}\right)g_{lm}^{-\text{ret}}Y_{lm}\right]+\dots\,,\nonumber 
\end{align}
and
\begin{align*}
F_{AB} & =\log r\sum_{l,m}\left(-1\right)^{l}h_{lm}^{-\text{ret}}\sqrt{\gamma}\left(\epsilon_{BC}\partial_{A}\partial^{C}Y_{lm}-\epsilon_{AC}\partial_{B}\partial^{C}Y_{lm}\right)\\
 & +\sum_{l,m}\left[\frac{d^{l}}{dv^{l}}h_{lm}^{\text{adv}}\left(v\right)+\left(-1\right)^{l}\left(h_{lm}^{-\text{ret}}\log2+H_{lm}^{-\text{ret}}\right)-\left(\sum_{k=0}^{l}\frac{\left(-1\right)^{k}}{\left(k!\right)^{2}}\frac{\left(l+k\right)!}{\left(l-k\right)!}H_{k}\right)h_{lm}^{-\text{ret}}\right]\\
&\times\sqrt{\gamma}\left(\epsilon_{BC}\partial_{A}\partial^{C}Y_{lm}-\epsilon_{AC}\partial_{B}\partial^{C}Y_{lm}\right)+\dots\,.
\end{align*}

\subsubsection{Matching conditions}

The matching conditions on the field strength follow immediately from the conditions relating the advanced and retarded fields obtained at $i^0$ in Eq.~\eqref{eq:match}. 
Indeed, from Eqs. \eqref{eq:Furlog} and \eqref{eq:Fvrlog}, the leading
logarithmic terms of $F_{ur}$ and $F_{vr}$ are given 
\[
\left.F_{ur}^{\text{log}}\left(\hat{x}\right)\right|_{\mathscr{I}_{-}^{+}}=\sum_{l,m}\left(-1\right)^{l}l\left(l+1\right)g_{lm}^{+\text{adv}}Y_{lm}\left(\hat{x}\right),
\]
\[
\left.F_{vr}^{\text{log}}\left(\hat{x}\right)\right|_{\mathscr{I}_{+}^{-}}=\sum_{l,m}\left(-1\right)^{l+1}l\left(l+1\right)g_{lm}^{-\text{ret}}Y_{lm}\left(\hat{x}\right).
\]
Therefore, using (\ref{eq:match}), we find the following matching
condition:
\[
\left.F_{vr}^{\text{log}}\left(-\hat{x}\right)\right|_{\mathscr{I}_{+}^{-}}=-\left.F_{ur}^{\text{log}}\left(\hat{x}\right)\right|_{\mathscr{I}_{-}^{+}}.
\]
Similarly, 
\begin{align*}
\left.F_{AB}^{\text{log}}\left(\hat{x}\right)\right|_{\mathscr{I}_{-}^{+}} & =\sum_{l,m}\left(-1\right)^{l}h_{lm}^{+\text{adv}}\sqrt{\gamma}\left(\epsilon_{BC}\partial_{A}\partial^{C}Y_{lm}-\epsilon_{AC}\partial_{B}\partial^{C}Y_{lm}\right)\,,
\end{align*}
and
\begin{align*}
\left.F_{AB}^{\text{log}}\left(\hat{x}\right)\right|_{\mathscr{I}_{+}^{-}} & =\sum_{l,m}\left(-1\right)^{l}h_{lm}^{-\text{ret}}\sqrt{\gamma}\left(\epsilon_{BC}\partial_{A}\partial^{C}Y_{lm}-\epsilon_{AC}\partial_{B}\partial^{C}Y_{lm}\right)\,.
\end{align*}
Thus,
\[
\left.F_{AB}^{\text{log}}\left(-\hat{x}\right)\right|_{\mathscr{I}_{+}^{-}}=\left.F_{AB}^{\text{log}}\left(\hat{x}\right)\right|_{\mathscr{I}_{+}^{-}}\,.
\]
In this case, we must take into account that the magnetic sector has
an extra minus sign in the parity coming from the Levi-Civita symbol,
i.e., $$\sqrt{\gamma}\epsilon_{AB}\partial^{B}Y_{lm}\left(-\hat{x}\right)=\left(-1\right)^{l+1}\sqrt{\gamma}\epsilon_{AB}\partial^{B}Y_{lm}\left(\hat{x}\right)$$
As shown in \cite{Fuentealba:2025ekj}, the matching conditions  for the coefficients of the leading (logarithmic) terms have opposite sign compared with those of \cite{He:2014cra,Kapec:2015ena,Strominger:2017zoo}.  If one assumes that the leading logarithmic terms are absent, the matching conditions for the coefficients of the next terms, which become leading, obey the matching conditions with the sign of \cite{He:2014cra,Kapec:2015ena,Strominger:2017zoo} characteristic of the $P$-branch.  The derivation proceeds as in the scalar case.  

\section{Conclusions}
\label{Sec:Conclusions}
In this work, we have studied the role of advanced and retarded radiation
in determining the asymptotic structure of massless scalar and electromagnetic
fields that exhibit radial logarithmic terms near $\mathscr{I}^{\pm}$.
Assuming that the total incoming energy flux across $\mathscr{I}^{-}$
is finite, we have shown that the final advanced wave, characterized
by an asymptotic $\log v$ dependence at late times, leaves a distinctive
imprint at $\mathscr{I}^{+}$ in the form of $\log r$ terms. Conversely,
the first retarded wave completely determines the radial logarithmic
behavior in the vicinity of $\mathscr{I}^{-}$. This analysis offers
both a novel perspective and a coherent physical interpretation of
the origin of the radial logarithmic terms. Furthermore, the results
obtained through this approach coincide with those in Refs. \cite{Fuentealba:2024lll,Fuentealba:2025ekj},
where a spatial-infinity-based route was used. 

Remarkably, this advanced/retarded wave approach enables the determination of the matching conditions for the radial logarithmic terms between the future of past null infinity and the past of future null infinity. This is achieved by evaluating the fields in the vicinity of $i^{0}$ and imposing that their asymptotic behaviour be of the Coulombic type, free of $\log r$ contributions at spatial infinity. The matching conditions derived using this approach are in exact agreement with those in Refs. \cite{Fuentealba:2024lll,Fuentealba:2025ekj}.  Our derivation not only confirms, but also provides a new physical insight on these matching conditions.

As we just stressed,  the assumption that there is no logarithmic term at spatial infinity plays a key role  in our derivation of the matching conditions.  This assumption is motivated by the asymptotic behaviour of the elementary solution of the Poissson equation (and of  superpositions of such solutions).  However, this assumption can be relaxed while preserving finiteness of Poincar\'e charges and of the energy fluxes through past and future infinity.  For instance $\log r/r$ terms with coefficients having definite parity at spatial infinity could be contemplated.  It is beyond the scope of this paper to investigate the most liberal asymptotic behaviour at spatial infinity that is consistent with the above physical requirements, since the assumed $1/r$ behaviour is sufficient for our purposes.

One might wonder whether the dominant logarithmic terms found at null infinity do not spoil convergence of the symplectic form and of the angular momentum.  While we have not directly studied this question at null infinity, the issue was indirectly addressed on Cauchy hypersurfaces (spatial infinity) in \cite{Henneaux:2018gfi,Henneaux:2018hdj,Henneaux:2018mgn}, where it was shown that no problem arises if only the dominant $P$-branch or the dominant $Q$-branch is present.  Indeed both branches decay with exactly the same powers of $r$ at spatial infinity, and the $Q$-branch does not exhibit $\log r$ term. Each branch is separately manifestly consistent on Cauchy hypersurfaces.  It is only when both are simultaneously present that divergences appear\footnote{In this respect, it is somewhat intriguing, and in any case consistent with finiteness,  that the (imaginary part of the) solution of \cite{Kim:2023qbl}, appendix D, has a pure dominant $Q$-term.}.   Since the ADM angular momentum at spatial infinity is finite for a pure dominant $Q$ branch, it provides a finite initial condition at null infinity ($u \rightarrow - \infty$) for the angular momentum even in that case, guaranteeing a finite flux as long as the angular momentum remains finite at later $u$'s.  It would be of interest to fully explore this question.  

It is natural to ask to what extent the results of this paper can be extended to the gravitational case. For linearized gravity, one may naturally expect a behaviour closely analogous to that of the scalar and electromagnetic fields discussed in this work. However, in the fully nonlinear theory, the situation is considerably more subtle. This is because, as observed for the scalar and electromagnetic fields discussed above, all subleading terms in the radial expansion near $\mathscr{I}^-$ contribute to the leading $\log r$ term at  $\mathscr{I}^+$. In General Relativity, it is extremely challenging to control all subleading terms in the radial expansion due to the inherent nonlinearities of the theory. However, since Bondi's energy-loss formula \cite{Bondi:1960jsa} has a similar form to that of the scalar and electromagnetic cases, the requirement of finite total energy implies that the shear tensor at late advanced times must take the form $C_{AB}\underset{v\rightarrow\infty}{=}C_{AB}^{+}\log v+\dots$. Consequently, when it is reexpressed in retarded coordinates, it will generically give rise to $\log r$  contributions in the asymptotic expansion near $\mathscr{I}^+$. How to connect these logarithmic terms to the polyhomogeneous solutions involving dominant logarithmic terms presented for example, in Refs. \cite{Winic:1985,Chrusciel:1993hx,Andersson:1994ng} remains an interesting question for future investigation.

\section*{Acknowledgements}
The research of AP is partially supported by Fondecyt grant
1220910. AP acknowledges support from the Erwin Schr\"odinger
International Institute for Mathematics and Physics (ESI) where some
of the research was undertaken during the ESI Research in Teams
Programme. The work of HG and MB is funded by FONDECYT grant 1230853. The work of MH is partially supported by FNRS-Belgium (convention IISN 4.4503.15), as well as by research funds from the Solvay Family.

\section*{Appendices}
\appendix

\section{General analysis of the massless scalar field}
\label{App:Gen}

\subsection{Retarded and advanced waves}

In this section, we extend the scalar field analysis of spherical and dipole waves 
to the general solution of the wave equation. As for the dipole case, subleading terms in the asymptotic radial expansion must be taken into account, as they play a fundamental role.

The general solution of the wave equation in a source-free region
is expressed again as the sum of retarded and advanced terms.
\[
\phi=\phi_{R}+\phi_{A},
\]
with
\begin{align}
\phi_{R} & =\sum_{l,m}\sum_{k=0}^{l}\frac{1}{2^{k}k!r^{k+1}}\frac{\left(l+k\right)!}{\left(l-k\right)!}\frac{d^{l-k}}{du^{l-k}}a_{l,m}\left(u\right)Y_{l m}\left(\theta,\phi\right),\label{eq:phi_R}\\
\phi_{A} & =\sum_{l,m}\sum_{k=0}^{l}\frac{\left(-1\right)^{k}}{2^{k}k!r^{k+1}}\frac{\left(l+k\right)!}{\left(l-k\right)!}\frac{d^{l-k}}{dv^{l-k}}b_{l,m}\left(v\right)Y_{l m}\left(\theta,\phi\right).\label{eq:phi_A}
\end{align}
Here, $a_{l,m}\left(u\right)$ and $b_{l,m}\left(v\right)$
are arbitrary functions of the retarded and advanced times, respectively.
It is noteworthy that the solution above is expressed purely
as a power-law expansion in the radial coordinate, with no logarithmic
terms.

In particular, the leading-order term at large $r$ takes the form:
\[
\phi_R =\frac{\bar{\phi}_{R}\left(u,\hat{x}\right)}{r}+O\left(r^{-2}\right) \, , \quad \phi_A = \frac{\bar{\phi}_{A}\left(v,\hat{x}\right)}{r}+O\left(r^{-2}\right) \, ,
\]
where 
\begin{align*}
\bar{\phi}_{R}\left(u,\hat{x}\right) & =\sum_{l,m}\frac{d^{l}}{du^{l}}a_{l,m}\left(u\right)Y_{l m},\\
\bar{\phi}_{A}\left(v,\hat{x}\right) & =\sum_{l,m}\frac{d^{l}}{dv^{l}}b_{l,m}\left(v\right)Y_{l m}.
\end{align*}
The total energy flux across $\mathscr{I}^{\pm}$ takes exactly the
same form as in Eqs. (\ref{eq:energy_ret}) and (\ref{eq:energy_adv}).
Consequently, for the $l$-th spherical harmonic, the requirement
of finite energy flux across $\mathscr{I}^{\pm}$ implies that  
\begin{align*}
\lim_{u\rightarrow\pm\infty}\frac{d^{l}}{du^{l}}a_{l,m}\left(u\right) & =\phi_{R\left(l,m\right)}^{\pm}\log\left(\pm u\right)+\varphi_{R\left(l,m\right)}^{\pm}+\dots\,,\\
\lim_{v\rightarrow\pm\infty}\frac{d^{l}}{dv^{l}}b_{l,m}\left(v\right) & =\phi_{A\left(l,m\right)}^{\pm}\log\left(\pm v\right)+\varphi_{A\left(l,m\right)}^{\pm}+\dots\, ,
\end{align*}
assuming again integer power law behaviour for $\partial_u \bar{\phi}_{R}$ ($\partial_v \bar{\phi}_{A}$).

Let us examine the contribution of the advanced wave to the asymptotic
behavior of the field near $\mathscr{I}^{+}$. This requires integrating
the following equation
\[
\frac{d^{l}}{dv^{l}}b_{l,m}\left(v\right)\underset{v\rightarrow+\infty}{=}\phi_{A\left(l,m\right)}^{+}\log\left(v\right)+\varphi_{A\left(l,m\right)}^{+}+\dots\,.
\]
To carry out the integration, we will make use of the following integral
identity 
\begin{equation}
\alpha\underbrace{\int\cdots\int}_{k}\log x\,dx+\beta\underbrace{\int\cdots\int}_{k}\,dx=\alpha\left(\frac{x^{k}}{k!}\log\left(x\right)-\frac{1}{k!}H_{k}x^{k}\right)+\beta\frac{x^{k}}{k!}+\sum_{n=0}^{k-1}c_{k-n}\frac{x^{n}}{n!}\,.\label{eq:ID}
\end{equation}
Here, $c_{k-n}$ denotes the integration constants, and $H_{k}=\sum_{n=1}^{k}\frac{1}{n}$
represents the harmonic numbers. 
This identity can be readily proven by induction, starting from its differential form,
\[
\frac{d^{k}}{dx^{k}}\left[\alpha\left(\frac{x^{k}}{k!}\log\left(x\right)-\frac{1}{k!}H_{k}x^{k}\right)+\beta\frac{x^{k}}{k!}+\sum_{n=0}^{k-1}c_{k-n}\frac{x^{n}}{n!}\right]=\alpha\log\left(x\right)+\beta\,,
\]
and using the recursive property of the harmonic numbers, $H_{k+1}=H_{k}+\frac{1}{k+1}$. 

Thus,
\begin{align}
\frac{d^{l-k}}{dv^{l-k}}b_{l,m}\left(v\right)\underset{v\rightarrow\infty}{=} & \phi_{A\left(l,m\right)}^{+}\left(\frac{v^{k}}{k!}\log\left(v\right)-\frac{1}{k!}H_{k}v^{k}\right)+\varphi_{A\left(l,m\right)}\frac{v^{k}}{k!}+\sum_{n=0}^{k-1}c_{A\left(k-n\right)}^{\left(l,m\right)}\frac{v^{n}}{n!}\,.\label{eq:db}
\end{align}
The term with $k=0$ in the above expression can be properly defined
by setting $H_{0}=0$. Substituting Eq.~\eqref{eq:db} into Eq.~\eqref{eq:phi_A}
one finds that the advanced solution at large $v$ can then be written
as
\begin{equation}
\phi_{A}\underset{v\rightarrow\infty}{=}\sum_{l,m}\sum_{k=0}^{l}\frac{\left(-1\right)^{k}}{2^{k}k!r^{k+1}}\frac{\left(l+k\right)!}{\left(l-k\right)!}\left[\phi_{A\left(l,m\right)}^{+}\left(\frac{v^{k}}{k!}\log\left(v\right)-\frac{1}{k!}H_{k}v^{k}\right)+\varphi_{A\left(l,m\right)}\frac{v^{k}}{k!}+\dots\right]Y_{l m}.\label{eq:Fieldv}
\end{equation}
We omit the terms containing the integration constants $c_{A(k-n)}^{(l,m)}$, as they contribute only subleading corrections that do not affect the present analysis.

The key step is to express the advanced coordinate in terms of the
retarded one using $v=u+2r$. This allows one to write
\begin{multline*}
\phi_{A}=\sum_{l,m}\sum_{k=0}^{l}\frac{\left(-1\right)^{k}}{2^{k}k!r^{k+1}}\frac{\left(l+k\right)!}{\left(l-k\right)!}\left[\phi_{A\left(l,m\right)}^{+}\left(\frac{\left(u+2r\right)^{k}}{k!}\log\left(u+2r\right)-\frac{1}{k!}H_{k}\left(u+2r\right)^{k}\right)\right.\\\left.+\varphi_{A\left(l,m\right)}\frac{\left(u+2r\right)^{k}}{k!}+\dots\right]Y_{l m}.
\end{multline*}
In the limit $r\rightarrow\infty$ with $u$ held constant, the leading
terms are given by
\begin{multline*}
\phi_{A} =\left[\sum_{l,m}\phi_{A\left(l,m\right)}^{+}\left(\sum_{k=0}^{l}\frac{\left(-1\right)^{k}}{\left(k!\right)^{2}}\frac{\left(l+k\right)!}{\left(l-k\right)!}\right)Y_{l m}\right]\frac{\log r}{r}\\ +\left[\sum_{l,m}\sum_{k=0}^{l}\frac{\left(-1\right)^{k}}{\left(k!\right)^{2}}\frac{\left(l+k\right)!}{\left(l-k\right)!}\left[\phi_{A\left(l,m\right)}^{+}\left(\log2-H_{k}\right)+\varphi_{A\left(l,m\right)}\right]Y_{l m}\right]\frac{1}{r}+\dots\,.
\end{multline*}
The above expression can be simplified using the identity \eqref{Plm1}
\[
\sum_{k=0}^{l}\frac{\left(-1\right)^{k}}{\left(k!\right)^{2}}\frac{\left(l+k\right)!}{\left(l-k\right)!}=\left(-1\right)^{l}\,.
\]
Consequently, if one considers the sum of the advanced and retarded
solutions, the full solution near $\mathscr{I}^{+}$ becomes:
\begin{align}
\left.\phi\right|_{\mathscr{I}^{+}} & =\left[\sum_{l,m}\left(-1\right)^{l}\phi_{A\left(l,m\right)}^{+}Y_{l m}\right]\frac{\log r}{r}\nonumber \\
 & +\left[\sum_{l,m}\left(\frac{d^{l}}{du^{l}}a_{l,m}\left(u\right)+\left(-1\right)^{l}\left(\phi_{A\left(l,m\right)}^{+}\log2+\varphi_{A\left(l,m\right)}\right)\right.\right.\label{eq:phi_I+}\\&-\left.\left.\left(\sum_{k=0}^{l}\frac{\left(-1\right)^{k}}{\left(k!\right)^{2}}\frac{\left(l+k\right)!}{\left(l-k\right)!}H_{k}\right)\phi_{A\left(l,m\right)}^{+}\right)Y_{l m}\right]\frac{1}{r}+\dots\,. \nonumber 
\end{align}
Note that a $\log r/r$ term appears, arising exclusively from the
advanced wave. This term is time-independent and typically includes
all spherical harmonics. Due to its time-independence, it does not
contribute to the energy flux across $\mathscr{I}^{+}$. Additionally,
the advanced wave also contributes a time-independent term at order
$r^{-1}$. 

Let us now determine the contribution of the retarded wave to the
asymptotic expansion near $\mathscr{I}^{-}$. The requirement of finite
total radiated energy implies that

\[
\frac{d^{l}}{du^{l}}a_{l,m}\left(u\right)\underset{u\rightarrow-\infty}{=}\phi_{R\left(l,m\right)}^{-}\log\left(-u\right)+\varphi_{R\left(l,m\right)}^{-}+\dots\,.
\]
Integrating the previous equation using the identity along the same lines of \eqref{eq:ID}
\[
\alpha\underbrace{\int\cdots\int}_{k}\log\left(-x\right)\,dx+\beta\underbrace{\int\cdots\int}_{k}\,dx=\alpha\left(\frac{x^{k}}{k!}\log\left(-x\right)-\frac{1}{k!}H_{k}x^{k}\right)+\beta\frac{x^{k}}{k!}+\sum_{n=0}^{k-1}c_{k-n}\frac{x^{n}}{n!}\,,
\]
one obtains
\begin{align*}
\frac{d^{l-k}}{du^{l-k}}a_{l,m}\left(u\right)\underset{v\rightarrow-\infty}{=} & \phi_{R\left(l,m\right)}^{-}\left(\frac{u^{k}}{k!}\log\left(-u\right)-\frac{1}{k!}H_{k}u^{k}\right)+\varphi_{R\left(l,m\right)}^{-}\frac{u^{k}}{k!}+\sum_{n=0}^{k-1}c_{R\left(k-n\right)}^{\left(l,m\right)}\frac{u^{n}}{n!}\,.
\end{align*}
Substituting this result into Eq.~\eqref{eq:phi_R}, one finds the
following expression for the retarded solution in the limit $u\rightarrow-\infty$:
\begin{equation}
\phi_{R}=\sum_{l,m}\sum_{k=0}^{l}\frac{1}{2^{k}k!r^{k+1}}\frac{\left(l+k\right)!}{\left(l-k\right)!}\left[\phi_{R\left(l,m\right)}^{-}\left(\frac{u^{k}}{k!}\log\left(-u\right)-\frac{1}{k!}H_{k}u^{k}\right)+\varphi_{R\left(l,m\right)}^{-}\frac{u^{k}}{k!}+\dots\right]Y_{l m}.\label{eq:Field_u}
\end{equation}
By expressing the retarded time $u$ in terms of the advanced time
$v$ via the relation $u=v-2r$, one finds
\begin{align*}
\phi_{R}=\sum_{l,m}\sum_{k=0}^{l}\frac{1}{2^{k}k!r^{k+1}}\frac{\left(l+k\right)!}{\left(l-k\right)!}\left[\phi_{R\left(l,m\right)}^{-}\left(\frac{\left(v-2r\right)^{k}}{k!}\log\left(2r-v\right)-\frac{1}{k!}H_{k}\left(v-2r\right)^{k}\right)\right. \\ \left.+\varphi_{R\left(l,m\right)}^{-}\frac{\left(v-2r\right)^{k}}{k!}+\dots\right]Y_{l m}.
\end{align*}
Therefore, in the limit $r\rightarrow\infty$ with $v$ held constant,
the leading terms are:
\begin{align*}
\phi_{R}=&\left[\sum_{l,m}\phi_{R\left(l,m\right)}^{-}\left(\sum_{k=0}^{l}\frac{\left(-1\right)^{k}}{\left(k!\right)^{2}}\frac{\left(l+k\right)!}{\left(l-k\right)!}\right)Y_{l m}\right]\frac{\log r}{r}\\
&+\left[\sum_{l,m}\sum_{k=0}^{l}\frac{\left(-1\right)^{k}}{\left(k!\right)^{2}}\frac{\left(l+k\right)!}{\left(l-k\right)!}\left[\phi_{R\left(l,m\right)}^{-}\left(\log2-H_{k}\right)+\varphi_{R\left(l,m\right)}^{-}\right]Y_{l m}\right]\frac{1}{r}+\cdots\,.
\end{align*}
Consequently, the full solution near $\mathscr{I}^{-}$ becomes
\begin{align}
\left.\phi\right|_{\mathscr{I}^{-}} & =\left[\sum_{l,m}\left(-1\right)^{l}\phi_{R\left(l,m\right)}^{-}Y_{l m}\right]\frac{\log r}{r}\nonumber \\
 & +\left[\sum_{l,m}\left(\left(\frac{d^{l}}{dv^{l}}b_{l,m}\left(v\right)\right)+\left(-1\right)^{l}\left(\phi_{R\left(l,m\right)}^{-}\log2+\varphi_{R\left(l,m\right)}^{-}\right)\right. \right. \label{eq:phi_I-}\\ & \left. \left.-\left(\sum_{k=0}^{l}\frac{\left(-1\right)^{k}}{\left(k!\right)^{2}}\frac{\left(l+k\right)!}{\left(l-k\right)!}H_{k}\right)\phi_{R\left(l,m\right)}^{-}Y_{l m}\right)\right]\frac{1}{r}+\cdots\,. \nonumber
\end{align}
In this case, there is a time-independent $\log r/r$ contribution
arising exclusively from the retarded wave.

The expressions (\ref{eq:phi_I+}) and (\ref{eq:phi_I-}), which describe
the behavior of the field near $\mathscr{I}^{\pm}$, match precisely
the asymptotic expansion introduced in Ref. \cite{Fuentealba:2024lll}
and given in Eqs. (\ref{eq:psi}) and (\ref{eq:psiprime}), where
\begin{align}
\Psi\left(\hat{x}\right) & =2\sum_{l,m}\left(-1\right)^{l}\phi_{A\left(l,m\right)}^{+}Y_{l m}\left(\hat{x}\right)\,,\label{eq:Psiell}\\
\Psi'\left(\hat{x}\right) & =2\sum_{l,m}\left(-1\right)^{l}\phi_{R\left(l,m\right)}^{-}Y_{l m}\left(\hat{x}\right)\,.\label{eq:psipell}
\end{align}
In the next subsection, we will show that this prescription directly
yields the matching conditions for the scalar field near spatial infinity.

\subsection{Matching conditions}

Let us examine the behavior of the field $\phi$ and its canonical
momentum $\pi=\partial_{t}\phi$ near spatial infinity $i^{0}$. 

For the advanced solution, one must replace $v=t+r$ in Eq. (\ref{eq:Fieldv})
and then take the limit $r\rightarrow\infty$ while keeping $t$ fixed.

\begin{align*}
\left.\phi_{A}\right|_{i^{0}} =
&\sum_{l,m}\sum_{k=0}^{l}
   \frac{(-1)^{k}}{2^{k}k!\,r^{k+1}}
   \frac{(l+k)!}{(l-k)!}
   \Bigg[
      \phi_{A(l,m)}^{+}
      \!\left(
         \frac{(t+r)^{k}}{k!}\log(t+r)
         -\frac{H_{k}}{k!}(t+r)^{k}
      \right)
     \\& + \varphi_{A(l,m)}\,\frac{(t+r)^{k}}{k!}
   \Bigg] Y_{l m}
   + \dots
\\
=
&\Bigg[
   \sum_{l,m}\phi_{A(l,m)}^{+}
   \!\left(
      \sum_{k=0}^{l}
         \frac{(-1)^{k}}{2^{k}(k!)^{2}}
         \frac{(l+k)!}{(l-k)!}
   \right)
   Y_{l m}
\Bigg]\frac{\log r}{r}
\\
&+
\Bigg[
   \sum_{l,m}\sum_{k=0}^{l}
      \frac{(-1)^{k}}{2^{k}(k!)^{2}}
      \frac{(l+k)!}{(l-k)!}
      \big(
         \varphi_{A(l,m)}
         - \phi_{A(l,m)}^{+} H_{k}
      \big)
      Y_{l m}
\Bigg]\frac{1}{r}
+ \dots\,.
\end{align*}
Similarly, for the retarded solution one replaces $u=t-r$ in Eq.
(\ref{eq:Field_u}) and then take the limit $r\rightarrow\infty$
while maintaining $t$ fixed.
\begin{align*}
\left.\phi_{R}\right|_{i^{0}}
&=
   \sum_{l,m, k\leq l }
   \frac{1}{2^{k}k!\,r^{k+1}}
   \frac{(l+k)!}{(l-k)!}
   \frac{(t-r)^{k}}{k!} \Bigg(
      \phi_{R(l,m)}^{-}
       \!\left(
         \log(r-t)
         -H_k
      \right)
     +\varphi_{R(l,m)}^{-}
   \Bigg)Y_{l m}
   +\dots
\\[4pt]
&=
\Bigg[
   \sum_{l,m}\phi_{R(l,m)}^{-}
   \!\left(
      \sum_{k=0}^{l}
         \frac{(-1)^{k}}{2^{k}(k!)^{2}}
         \frac{(l+k)!}{(l-k)!}
   \right)
   Y_{l m}
\Bigg]\frac{\log r}{r}
\\[4pt]
&\quad+
\Bigg[
   \sum_{l,m, k\leq l }
      \frac{(-1)^{k}}{2^{k}(k!)^{2}}
      \frac{(l+k)!}{(l-k)!}
      \big(
         \varphi_{R(l,m)}^{-}
         -\phi_{R(l,m)}^{-}H_{k}
      \big)
      Y_{l m}
\Bigg]\frac{1}{r}
+\dots\,.
\end{align*}
The above expressions can be simplified using the identity \eqref{Pl0}
\[
\sum_{k=0}^{l}\frac{\left(-1\right)^{k}}{2^{k}\left(k!\right)^{2}}\frac{\left(l+k\right)!}{\left(l-k\right)!}=P_l(0).
\]
The complete expression near $i^{0}$ is thus the sum of the advanced
and retarded terms, which gives
\begin{align}
\left.\phi\right|_{i^{0}} & =\left[\sum_{l,m}P_{l}(0)\left(\phi_{A\left(l,m\right)}^{+}+\phi_{R\left(l,m\right)}^{-}\right)Y_{l m}\right]\frac{\log r}{r}+\nonumber \\
 & +\left[\sum_{l,m}\Bigg(P_{l}(0)\left(\varphi_{A\left(l,m\right)}+\varphi_{R\left(l,m\right)}^{-}\right)\right.
 \label{eq:phii0} \\
 &\hspace{1cm}-\left.\sum_{k=0}^{l}\frac{\left(-1\right)^{k}}{2^{k}\left(k!\right)^{2}}\frac{\left(l+k\right)!}{\left(l-k\right)!}H_{k}\left(\phi_{A\left(l,m\right)}^{+}+\phi_{R\left(l,m\right)}^{-}\right)\Bigg)Y_{l m}\right]\frac{1}{r}+\dots\,.  \nonumber
\end{align}

Let us now compute the canonical momentum $\pi=\partial_{t}\phi$
near $i^{0}$. The contribution from the advanced solution is given
by
\begin{align*}
\left.\pi_{A}\right|_{i^{0}} &=\left[\sum_{l,m}\left(\sum_{k=0}^{l}\frac{\left(-1\right)^{k}k}{2^{k}\left(k!\right)^{2}}\frac{\left(l+k\right)!}{\left(l-k\right)!}\right)\phi_{A\left(l,m\right)}^{+}Y_{l m}\right]\frac{\log r}{r^{2}}\\
 & +\left[\sum_{l,m}\left(\sum_{k=0}^{l}\frac{\left(-1\right)^{k}}{2^{k}\left(k!\right)^{2}}\frac{\left(l+k\right)!}{\left(l-k\right)!}\left(\phi_{A\left(l,m\right)}^{+}\left(1-kH_{k}\right)+k\varphi_{A\left(l,m\right)}\right)\right)Y_{l m}\right]\frac{1}{r^{2}}+\dots\,.
\end{align*}
The retarded component of the momentum near $i^{0}$ takes the form
\begin{align*}
\left.\pi_{R}\right|_{i^{0}}= & -\left[\sum_{l,m}\sum_{k=1}^{l}\frac{\left(-1\right)^{k}k}{2^{k}\left(k!\right)^{2}}\frac{\left(l+k\right)!}{\left(l-k\right)!}\phi_{R\left(l,m\right)}^{-}Y_{l m}\right]\frac{\log r}{r^{2}}+\\
 & +\left[\sum_{l,m}\sum_{k=0}^{l}\frac{\left(-1\right)^{k}}{2^{k}\left(k!\right)^{2}}\frac{\left(l+k\right)!}{\left(l-k\right)!}\left(\phi_{R\left(l,m\right)}^{-}\left(1+kH_{k}\right)-k\varphi_{R\left(l,m\right)}^{-}\right)Y_{l m}\right]\frac{1}{r^{2}}+\dots\,.
\end{align*}
This expression can be simplified by applying the following identity \eqref{Pl10}
\[
\sum_{k=0}^{l}\frac{\left(-1\right)^{k}k}{2^{k}\left(k!\right)^{2}}\frac{\left(l+k\right)!}{\left(l-k\right)!}=-P_l'(0),
\]
Thus, the combined contribution of the retarded and advanced components
of the momentum near $i^{0}$ becomes 
\begin{align}
\left.\pi\right|_{i^{0}} & =-\left[\sum_{l,m}P_l'(0)\left(\phi_{A\left(l,m\right)}^{+}-\phi_{R\left(l,m\right)}^{-}\right)Y_{l m}\right]\frac{\log r}{r^{2}}\nonumber \\
 & +\left[\sum_{l,m}\left(P_l(0)\left(\phi_{A\left(l,m\right)}^{+}+\phi_{R\left(l,m\right)}^{-}\right)-P'_{l}(0)\left(\varphi_{A\left(l,m\right)}-\varphi_{R\left(l,m\right)}^{-}\right)\right.\right.\label{eq:pii0}\\
 & +\left.\left.\left(\sum_{k=0}^{l}\frac{\left(-1\right)^{k}}{2^{k}\left(k!\right)^{2}}\frac{\left(l+k\right)!}{\left(l-k\right)!}kH_{k}\right)\left(\phi_{R\left(l,m\right)}^{-}-\phi_{A\left(l,m\right)}^{+}\right)\right)Y_{l m}\right]\frac{1}{r^{2}}+\dots\,
 \nonumber
\end{align}

As noted in \cite{Fuentealba:2024lll}, terms of order $\log r/r$
in $\left.\phi\right|_{i^{0}}$ and $\log r/r^{2}$ in $\left.\pi\right|_{i^{0}}$
are not allowed. Therefore, from Eqs. (\ref{eq:phii0}) and (\ref{eq:pii0}),
we obtain the following conditions
\begin{align}
\sum_{l,m}P_l(0)\left(\phi_{A\left(l,m\right)}^{+}+\phi_{R\left(l,m\right)}^{-}\right)Y_{l m} & =0\,,\label{eq:cond1}\\
\sum_{l,m}P'_l(0)\left(\phi_{A\left(l,m\right)}^{+}-\phi_{R\left(l,m\right)}^{-}\right)Y_{l m} & =0\,.\label{eq:cond2}
\end{align}
There are two cases depending if $l$ is odd or even. Thanks to the parity properties of the Legendre polynomials, we have that half of the equations are always satisfied. So, for $l$ even, only the first equation imposes a condition on the field coefficients. In contrast, for $l$ odd only the second relation yields a condition on the difference of the coefficients $\phi^\pm_{(l,m)}$. 

Consequently, the
condition of not having logarithmic terms in (\ref{eq:phii0}) and
(\ref{eq:pii0}) imposes the following relations between the coefficients
$\phi_{A\left(l,m\right)}^{+}$ and $\phi_{R\left(l,m\right)}^{-}$ that can be summarized as 
\begin{equation}
\phi_{A\left(l,m\right)}^{+}=-\left(-1\right)^{l}\phi_{R\left(l,m\right)}^{-}\,.\label{eq:phiAR}
\end{equation}
The above expression establishes a relation between the leading logarithmic
contributions of the field at $\mathscr{I}_{-}^{+}$ and $\mathscr{I}_{+}^{-}$.
Using the definitions of the radial logarithmic terms given in Eqs.(\ref{eq:Psiell})
and (\ref{eq:psipell}), and imposing the condition in (\ref{eq:phiAR}),
one obtains:

\begin{align*}
\Psi\left(\hat{x}\right)+\Psi'\left(-\hat{x}\right) & =2\sum_{l,m}\left(-1\right)^{l}\left(\phi_{A\left(l,m\right)}^{+}Y_{l m}\left(\hat{x}\right)+\phi_{R\left(l,m\right)}^{-}Y_{l m}\left(-\hat{x}\right)\right)\,,\\
 & =2\sum_{l,m}\left(-1\right)^{l}\left(\phi_{A\left(l,m\right)}^{+}+\left(-1\right)^{l}\phi_{R\left(l,m\right)}^{-}\right)Y_{l m}\left(\hat{x}\right)\,,\\
 & =0\,.
\end{align*}
Here, we have used the parity property of the spherical harmonics,
$Y_{l m}\left(-\hat{x}\right)=\left(-1\right)^{l}Y_{l m}\left(\hat{x}\right)$.
As a result, we obtain the relation

\[
\Psi\left(\hat{x}\right)=-\Psi'\left(-\hat{x}\right),
\]
which corresponds precisely to the matching condition established
in \cite{Fuentealba:2024lll}.

If we now assume that there is no leading logarithmic term ($\Psi(\hat x)= \Psi'(\hat x) = 0$), the solution reduces at leading order to the $P$-branch (in the limit $u \rightarrow - \infty$, $v \rightarrow \infty$), which obeys the ``standard'' parity condition of \cite{Campiglia:2017dpg}.  

\section{General solution of Maxwell's equations in retarded coordinates}
\label{App:Max}

The wave equation obeyed by the electromagnetic potential is
\[
\left(\boxempty A_{\mu}-\partial_{\mu}\nabla_{\nu}A^{\nu}\right)=0.
\]
Writing this solution in terms of retarded coordinates, and imposing
for simplicity the gauge $A_{r}=0$, one finds the following equations:
\begin{align}
\partial_{r}^{2}A_{u}+\frac{2}{r}\partial_{r}A_{u}-\partial_{u}\partial_{r}A_{u}+\frac{1}{r^{2}}\left(D^{2}A_{u}-\partial_{u}\left(D^{B}A_{B}\right)\right)&=0\,,\label{eq:equ-1}\\
\partial_{r}^{2}A_{u}+\frac{2}{r}\partial_{r}A_{u}-\frac{1}{r^{2}}\partial_{r}\left(D^{B}A_{B}\right)&=0\,,\label{eq:eqr-1}\\
\partial_{r}^{2}A_{A}-2\partial_{u}\partial_{r}A_{A}+\partial_{A}\partial_{r}A_{u}+\frac{D^{B}D_{B}A_{A}-D_{A}D_{C}A^{C}-A_{A}}{r^{2}}&=0\,.\label{eq:eqA-1}
\end{align}
To solve these equations, it is convenient to introduce the following separation of variables
\begin{align*}
A_{u} & =f_{lm}\left(u,r\right)Y_{lm}\,,\\
A_{A} & =g_{lm}\left(u,r\right)\partial_{A}Y_{lm}+h_{lm}\left(u,r\right)\sqrt{\gamma}\epsilon_{AB}\partial^{B}Y_{lm}\,.
\end{align*}
Thus, from Eq. \eqref{eq:equ-1} one finds
\[
\partial_{r}^{2}f_{lm}+\frac{2}{r}\partial_{r}f_{lm}-\partial_{u}\partial_{r}f_{lm}+\frac{1}{r^{2}}l\left(l+1\right)\left(\partial_{u}g_{lm}-f_{lm}\right)=0.
\]
From Eq. \eqref{eq:eqr-1} one obtains
\[
\partial_{r}^{2}f_{lm}+\frac{2}{r}\partial_{r}f_{lm}+\frac{1}{r^{2}}l\left(l+1\right)\partial_{r}g_{lm}=0\,.
\]
From Eq. \eqref{eq:eqA-1} one finds
\begin{multline*}
\left(\partial_{r}^{2}g_{lm}+\partial_{r}f_{lm}-2\partial_{u}\partial_{r}g_{lm}+\frac{1}{r^{2}}l\left(l+1\right)g_{lm}-\frac{1}{r^{2}}g_{lm}\right)\partial_{A}Y_{lm}\\+\left(\partial_{r}^{2}h_{lm}-2\partial_{u}\partial_{r}h_{lm}-\frac{1}{r^{2}}h_{lm}\right)\sqrt{\gamma}\epsilon_{AB}\partial^{B}Y_{lm}\\
  +\frac{1}{r^{2}}g_{lm}D^{C}D_{A}D_{C}Y_{lm}+\frac{1}{r^{2}}h_{lm}\sqrt{\gamma}\epsilon_{AB}D^{C}D^{B}D_{C}Y_{lm}=0.
\end{multline*}
Using the identity $D^{B}D_{A}D_{B}\Lambda=D_{A}D^{2}\Lambda+D_{A}\Lambda$,
one obtains
\begin{multline}
\left(\partial_{r}^{2}g_{lm}+\partial_{r}f_{lm}-2\partial_{u}\partial_{r}g_{lm}\right)\partial_{A}Y_{lm}+\\ \left(\partial_{r}^{2}h_{lm}-2\partial_{u}\partial_{r}h_{lm}-\frac{l\left(l+1\right)}{r^{2}}h_{lm}\right)\sqrt{\gamma}\epsilon_{AB}\partial^{B}Y_{lm}=0\,.
\end{multline}
This gives two equations
\[
0=\partial_{r}^{2}g_{lm}+\partial_{r}f_{lm}-2\partial_{u}\partial_{r}g_{lm}\,,
\]
\[
0=\partial_{r}^{2}h_{lm}-2\partial_{u}\partial_{r}h_{lm}-\frac{l\left(l+1\right)}{r^{2}}h_{lm}\,.
\]

In summary, Maxwell equations reduce to the following system of differential equations

\emph{Equations for $f$ and $g$}:
\begin{align}
0 & =\partial_{r}^{2}f_{lm}+\frac{2}{r}\partial_{r}f_{lm}-\partial_{u}\partial_{r}f_{lm}+\frac{1}{r^{2}}l\left(l+1\right)\left(\partial_{u}g_{lm}-f_{lm}\right)\,,\label{eq:f1}\\
0 & =\partial_{r}^{2}f_{lm}+\frac{2}{r}\partial_{r}f_{lm}+\frac{1}{r^{2}}l\left(l+1\right)\partial_{r}g_{lm}\,,\label{eq:f2}\\
0 & =\partial_{r}^{2}g_{lm}+\partial_{r}f_{lm}-2\partial_{u}\partial_{r}g_{lm}\,.\label{eq:f3}
\end{align}

\emph{Equation for $h$}:
\begin{equation}
0=\partial_{r}^{2}h_{lm}-2\partial_{u}\partial_{r}h_{lm}-\frac{l\left(l+1\right)}{r^{2}}h_{lm}\,.\label{eq:h1}
\end{equation}

We shall assume, as in the scalar case, that the functions $f_{lm}$, $g_{lm}$ and $h_{lm}$, can be  decomposed as the sum of  retarded and advanced waves.  
The retarded part is assumed to have an asymptotic expansion in inverse powers of $r$  with coefficients that are smooth functions of $u$ and of the angles.  Similarly, the advanced part is assumed to have an asymptotic expansion in inverse powers of $r$ with coefficients that are smooth functions of $v$ and of the angles. 

\subsection{Solution of the magnetic sector}

By performing the change of variables $h_{lm}\left(u,r\right)=r\phi\left(u,r\right)$
in Eq. (\ref{eq:h1}), the differential equation reduces to that of
a scalar field (Eq. (\ref{Eq:uWaveEq})). Therefore, the general solution to Eq. (\ref{eq:h1})
is given by 
\begin{equation}
h_{lm}\left(u,r\right)=\sum_{k=0}^{l}\frac{1}{k!\left(2r\right)^{k}}\frac{\left(l+k\right)!}{\left(l-k\right)!}\left(\frac{d^{l-k}}{du^{l-k}}h_{lm}^{\text{ret}}\left(u\right)+\left(-1\right)^{k}\frac{d^{l-k}}{dv^{l-k}}h_{lm}^{\text{adv}}\left(v\right)\right)\,.\label{eq:hur}
\end{equation}

\subsection{Solution of the electric sector}

The equations that must be solved are given by the following system
of differential equations:
\begin{align}
0 & =\partial_{r}^{2}f_{lm}+\frac{2}{r}\partial_{r}f_{lm}-\partial_{u}\partial_{r}f_{lm}+\frac{1}{r^{2}}l\left(l+1\right)\left(\partial_{u}g_{lm}-f_{lm}\right)\,,\label{eq:f1-2}\\
0 & =\partial_{r}^{2}f_{lm}+\frac{2}{r}\partial_{r}f_{lm}+\frac{1}{r^{2}}l\left(l+1\right)\partial_{r}g_{lm}\,,\label{eq:f2-2}\\
0 & =\partial_{r}^{2}g_{lm}+\partial_{r}f_{lm}-2\partial_{u}\partial_{r}g_{lm}\,.\label{eq:f3-2}
\end{align}
If we multiply (\ref{eq:f2-2}) by $r^{2}$ we find an exact differential
that integrates as follows:
\begin{equation}
r^{2}\partial_{r}f_{lm}+l\left(l+1\right)g_{lm}=\alpha_{lm}\left(u\right)l\left(l+1\right).\label{eq:exdiff}
\end{equation}
Replacing (\ref{eq:exdiff}) in (\ref{eq:f3-2}) we find the following
equation for $g_{lm}$:
\[
\partial_{r}^{2}g_{lm}-2\partial_{u}\partial_{r}g_{lm}-\frac{l\left(l+1\right)}{r^{2}}g_{lm}=-\frac{l\left(l+1\right)\alpha_{lm}\left(u\right)}{r^{2}}.
\]
Note that this equation is identical to that of the magnetic sector
in (\ref{eq:h1}) but with a source given by $\alpha_{lm}\left(u\right)$.
Therefore, the homogeneous solution will be identical to the one of
the magnetic sector in (\ref{eq:hur}). On the other hand, a particular
solution is given by $g_{lm}^{\text{part}}=\alpha_{lm}$. Consequently,
the general solution is given by 
\begin{equation}
g_{lm}\left(u,r\right)=\sum_{k=0}^{l}\frac{1}{2^{k}k!r^{k}}\frac{\left(l+k\right)!}{\left(l-k\right)!}\left(-\frac{d^{l-k}}{du^{l-k}}g_{lm}^{\text{ret}}\left(u\right)+\left(-1\right)^{k}\frac{d^{l-k}}{dv^{l-k}}g_{lm}^{\text{adv}}\left(v\right)\right)+\alpha_{lm}\left(u\right)\,,\label{eq:glm}
\end{equation}
for $l\geq1$. Note that the zero mode is irrelevant because it enters in the term
with $\partial_{A}Y_{lm}$. For future convenience, we have added
a minus sign in front of $g_{lm}^{R}\left(u\right)$.

In order to determine $f_{lm}\left(u,r\right)$, let us substract
equations (\ref{eq:f1-2}) and (\ref{eq:f2-2})
\[
0=-\partial_{u}\partial_{r}f_{lm}+\frac{1}{r^{2}}l\left(l+1\right)\left[\partial_{u}g_{lm}-f_{lm}-\partial_{r}g_{lm}\right].
\]
Taking a $u$-derivative of Eq. (\ref{eq:exdiff}) and replacing it
in the previous equation, we find
\[
0=\frac{1}{r^{2}}l\left(l+1\right)\left[2\partial_{u}g_{lm}-\partial_{r}g_{lm}-\partial_{u}\alpha_{lm}-f_{lm}\right].
\]
Note that for $l\geq1$, one obtains the following expression for $f_{lm}$
\[
f_{lm}\left(u,r\right)=2\partial_{u}g_{lm}\left(u,r\right)-\partial_{r}g_{lm}\left(u,r\right)-\partial_{u}\alpha_{lm}\left(u\right)\qquad\qquad\text{for }l\geq1.
\]
The case $l=0$ must be treated separately. Indeed, for $l=0$, the
equations (\ref{eq:f1-2})-(\ref{eq:f3-2}) reduce to 
\begin{align}
0 & =\partial_{r}^{2}f_{00}+\frac{2}{r}\partial_{r}f_{00}-\partial_{u}\partial_{r}f_{00}\,,\label{eq:f1-2-1}\\
0 & =\partial_{r}^{2}f_{00}+\frac{2}{r}\partial_{r}f_{00}\,,\label{eq:f2-2-1}\\
0 & =\partial_{r}^{2}g_{00}+\partial_{r}f_{00}-2\partial_{u}\partial_{r}g_{00}\,.\label{eq:f3-2-1}
\end{align}
In this case, the equations for $f_{lm}$ decouple automatically.
From Eqs. (\ref{eq:f1-2-1}) and (\ref{eq:f2-2-1}) we find
\[
f_{00}=\frac{Q}{r}-\partial_{u}\alpha_{00}\left(u\right)\,.
\]
 The zero mode $g_{00}$ can be determined from Eq. (\ref{eq:f3-2-1}),
but is completely irrelevant for the analysis because it appears in
the term with $\partial_{A}Y_{lm}$.

Consequently, the expression for $f_{lm}\left(u,r\right)$ is the
following:
\begin{align*}
f_{lm}\left(u,r\right) & =-\sum_{k=0}^{l}\frac{1}{2^{k-1}k!r^{k}}\frac{\left(l+k\right)!}{\left(l-k\right)!}\frac{d^{l-k+1}}{du^{l-k+1}}g_{lm}^{\text{ret}}\left(u\right)\\
 & +\sum_{k=0}^{l}\frac{k}{2^{k}k!r^{k+1}}\frac{\left(l+k\right)!}{\left(l-k\right)!}\left(-\frac{d^{l-k}}{du^{l-k}}g_{lm}^{\text{ret}}\left(u\right)+\left(-1\right)^{k}\frac{d^{l-k}}{dv^{l-k}}g_{lm}^{\text{adv}}\left(v\right)\right)+\partial_{u}\alpha_{lm}\left(u\right)\,.
\end{align*}
The expression for the electromagnetic potential is then given by
\begin{align*}
A_{u} & =-\sum_{l,m}\sum_{k=0}^{l}\frac{1}{2^{k-1}k!r^{k}}\frac{\left(l+k\right)!}{\left(l-k\right)!}\frac{d^{l-k+1}}{du^{l-k+1}}g_{lm}^{\text{ret}}\left(u\right) Y_{lm}\\
 & +\sum_{l,m}\sum_{k=0}^{l}\frac{k}{2^{k}k!r^{k+1}}\frac{\left(l+k\right)!}{\left(l-k\right)!}\left(-\frac{d^{l-k}}{du^{l-k}}g_{lm}^{\text{ret}}\left(u\right)+\left(-1\right)^{k}\frac{d^{l-k}}{dv^{l-k}}g_{lm}^{\text{adv}}\left(v\right)\right)Y_{lm}+\frac{Q}{r}+\partial_{u}\alpha\,,
\end{align*}
\begin{align*}
A_{A} & =\sum_{l,m}\sum_{k=0}^{l}\frac{1}{2^{k}k!r^{k}}\frac{\left(l+k\right)!}{\left(l-k\right)!}\left(-\frac{d^{l-k}}{du^{l-k}}g_{lm}^{\text{ret}}\left(u\right)+\left(-1\right)^{k}\frac{d^{l-k}}{dv^{l-k}}g_{lm}^{\text{adv}}\left(v\right)\right)\partial_{A} Y_{lm}+\partial_{A}\alpha\\
 & +\sum_{l,m}\sum_{k=0}^{l}\frac{1}{2^{k}k!r^{k}}\frac{\left(l+k\right)!}{\left(l-k\right)!}\left(\frac{d^{l-k}}{du^{l-k}}h_{lm}^{\text{ret}}\left(u\right)+\left(-1\right)^{k}\frac{d^{l-k}}{dv^{l-k}}h_{lm}^{\text{adv}}\left(v\right)\right)\sqrt{\gamma}\epsilon_{AB}\partial^{B}Y_{lm},
\end{align*}
where $\alpha$ is function of $u$ and $\hat{x}$.

It is convenient to fix the gauge such that near $\mathscr{I}^{+}$
one has $A_{u}=O\left(r^{-1}\right)$. This can be done by choosing
\[
\alpha\left(u,\hat{x}\right)=\sum_{l,m}2\frac{d^{l}}{du^{l}}g_{lm}^{\text{ret}}\left(u\right)Y_{lm}\left(\hat{x}\right)\,,
\]
and is compatible with $A_r= 0$ since $\alpha$ does not depend on $r$.  

With this gauge choice, we finally get for the full components (retarded + advanced) of the electromagnetic potential in retarded coordinates the following expressions,
\begin{align*}
A_{u} & =-\sum_{l,m}\sum_{k=1}^{l}\frac{1}{2^{k-1}k!r^{k}}\frac{\left(l+k\right)!}{\left(l-k\right)!}\frac{d^{l-k+1}}{du^{l-k+1}}g_{lm}^{\text{ret}}\left(u\right)Y_{lm}\\
 & +\sum_{l,m}\sum_{k=0}^{l}\frac{k}{2^{k}k!r^{k+1}}\frac{\left(l+k\right)!}{\left(l-k\right)!}\left(-\frac{d^{l-k}}{du^{l-k}}g_{lm}^{\text{ret}}\left(u\right)+\left(-1\right)^{k}\frac{d^{l-k}}{dv^{l-k}}g_{lm}^{\text{adv}}\left(v\right)\right)Y_{lm}+\frac{Q}{r}\,.
\end{align*}
\begin{align*}
A_{A} & =\sum_{l,m}\sum_{k=0}^{l}\frac{1}{2^{k}k!r^{k}}\frac{\left(l+k\right)!}{\left(l-k\right)!}\left(-\frac{d^{l-k}}{du^{l-k}}g_{lm}^{\text{ret}}\left(u\right)+\left(-1\right)^{k}\frac{d^{l-k}}{dv^{l-k}}g_{lm}^{\text{adv}}\left(v\right)\right)\partial_{A} Y_{lm}\\
 &+2\sum_{l,m}\frac{d^{l}}{du^{l}}g_{lm}^{\text{ret}}\left(u\right)\partial_{A} Y_{lm}\\
 & +\sum_{lm}\sum_{k=0}^{l}\frac{1}{2^{k}k!r^{k}}\frac{\left(l+k\right)!}{\left(l-k\right)!}\left(\frac{d^{l-k}}{du^{l-k}}h_{lm}^{\text{ret}}\left(u\right)+\left(-1\right)^{k}\frac{d^{l-k}}{dv^{l-k}}h_{lm}^{\text{adv}}\left(v\right)\right)\sqrt{\gamma}\epsilon_{AB}\partial^{B}Y_{lm}\,.
\end{align*}

\section{Legendre polynomials properties}
\label{LegendreAppendix}
In this article, multiple sums appeared, some of which have known results. In this section, we will show how to compute them.

There are two types of sums: those appearing at null infinity, which yield the matching conditions, and those arising when evaluating the field at spatial infinity. Both can be expressed in terms of Legendre polynomials and their derivatives, evaluated at specific points, either $x=-1$ or $x=0$.

\subsection{Values at $x=0$}
An interesting representation of the Legendre polynomials is given by
\begin{align}
    P_l(x)=\sum_{k=0}^{l}\frac{(x-1)^k}{2^k (k!)^2}\frac{(l+k)!}{(l-k)!}; \quad \text{with}\quad P_l(x)=(-1)^l P_l(-x). \label{legendresum}
\end{align}
This representation is useful because several of the relevant sums can be written as Legendre polynomials and their derivatives evaluated at zero:
\begin{align}
    P_l(0)=&\sum_{k=0}^{l} \frac{(-1)^k}{2^k (k!)^2} \frac{(l+k)!}{(l-k)!}; \quad P_l(0)=0 \text{ for }l\text{ odd}\label{Pl0}\\
    -P'_l(0)=&\sum_{k=0}^l \frac{k(-1)^k}{2^k(k!)^2} \frac{(l+k)!}{(l-k)!}\quad P'_l(0)=0 \text{ for }l\text{ even} \label{Pl10}\\
    P''_l(0)=&\sum_{k=0}^l \frac{k(k-1)(-1)^k}{2^k(k!)^2} \frac{(l+k)!}{(l-k)!}\quad P''_l(0)=0 \text{ for }l\text{ odd}.\label{Pl20}
\end{align}
Each of these sums arises when evaluating the fields at $i^0$. The appearance of these polynomials is not surprising, as they naturally occur in \cite{Fuentealba:2024lll,Fuentealba:2025ekj}.

\subsection{Values at $x=\pm1$}
The other type of sums in this article are those related to the field evaluated at null infinity (either future or past). In general, they can always be expressed as Legendre polynomials evaluated at $x=-1$. From the same representation in \eqref{legendresum}, we can write
\begin{align}
    P_l(-1)=&\sum_{k=0}^l \frac{(-1)^k}{(k!)^2}\frac{(l+k)!}{(l-k)!}\label{Plm1}\\
    -2P'_l(-1)=& \sum_{k=0}^l \frac{k(-1)^k}{(k!)^2}\frac{(l+k)!}{(l-k)!}.\label{Plm2}
\end{align}
Due to the parity condition and the representation of the Legendre polynomials in \eqref{legendresum} we can easily determine the value of $P_{l}(-1)$ and their derivatives 
\begin{align}
    P_{l}(-1)=(-1)^l P_l(1)=(-1)^l\,,\quad 
    -2P'_{l}(-1)=2(-1)^l P'_l(1)=(-1)^l l(l+1).
\end{align}

\providecommand{\href}[2]{#2}\begingroup\raggedright\endgroup


\end{document}